\documentstyle[12pt,epsf]{article}
\newcommand{\req}[1]{(\ref{#1})}
\newcommand{\bel}[1]{\begin{equation}\label{#1}}

\begin{document}
\def\kp{k^{\prime}}
\def\akp{a_k^\dagger}
\def\ajp{a_j^\dagger}
\def\ajb{a_{\bar j}}
\def\ajbp{a_{\bar j}^\dagger}
\def\akbp{a^\dagger_{\bar k}}
\def\akb{a_{\bar k}}
\def\ksp{js^{\pr}}
\def\uksp{u_{js^{\pr}}}
\def\vksp{v_{js^{\pr}}}
\def\sumkkp{\sum_{ksjs^{\pr}}}
\def\Deleq{{\Delta^{\rm eq}}}
\def\eps{\epsilon}
\def\epsk{\epsilon_k}
\def\epsj{\epsilon_j}
\def\alkp{\alpha_k^\dagger}
\def\alk{\alpha_k}
\def\alkbp{\alpha^\dagger_{\bar k}}
\def\alkb{\alpha_{\bar k}}
\def\aljp{\alpha_j^\dagger}
\def\alj{\alpha_j}
\def\aljbp{\alpha^\dagger_{\bar j}}
\def\aljb{\alpha_{\bar j}}
\def\gav{\gamma_{\rm av}}
\def\ep{e^{\prime}}
\def\om{\omega }
\def\Om{\Omega }
\def\pr{\prime}
\def\fkj{F_{kj}}
\def\tbhp{{\hat H}^{{\cal P}}}
\def\tbhpP{{\hat H}^{{\cal P\;\prime}}}
\def\fqpt{f(Q,P,t)}
\def\xvi{{\hat x}_i}
\def\pvi{{\hat p}_i}
\def\hbo{\hbar \om}
\def\zerofri{\gamma(0)}
\def\h#1{{\hat #1}}
\def\drft{{\tilde \chi}^{\prime \prime }}
\def\drf{\chi^{\prime \prime }}
\def\rft{{\tilde \chi}}
\def\rf#1{\chi (#1)}
\def\chiosc{\chi_{\rm osc}(\om)}
\def\bra{\big\langle}
\def\ket{\big\rangle}
\def\fmb{ \langle \hat F \rangle }
\def\ham#1{\hat H(\xvi,\pvi,#1)}
\def\hamrmf#1{\hat H_{\rm rmf}(\xvi,\pvi,#1)}
\def\hamsm#1{\hat H_{\rm sm}(\xvi,\pvi,#1)}
\def\ffield#1{ \hat F(\xvi,\pvi,#1)}
\def\equiop{\hat{\rho}_{\rm qs}}
\def\mev{\;{\rm MeV}}
\def\Vres{\hat V^{(2)}_{\rm res}(\xvi ,\pvi )}
\def\dtwohqzero{\left<{\partial^2\hat H\over \partial Q^2}(Q_0)
    \right>^{\rm qs}_{Q_0,T_0}}
\def\fext{f_{\rm ext}}
\def\respc{\chi_{\rm coll}(\om)}
\def\dpr{{\prime \prime}}
\def\corrheat{_0\psi^{\dpr }}
\def\self{{\it\Sigma}}
\def\selfre{\Sigma^\prime}
\def\foper{\hat F}
\vfill
\title{Pairing and shell effects in the transport
coefficients of collective motion
\thanks{Supported in part by the Deutsche Forschungsgemeinschaft}}
\author{F.A.Ivanyuk$^{1,2}$ and H.Hofmann$^{2}$\\
\small\it{1) Institute for Nuclear Research, Kiev-28, Ukraine, e-mail:
ivanyuk@kinr.kiev.ua}
\\
\small\it{2) Physik-Department der TUM, D-85747 Garching, Germany,}
\\\small\it{e-mail: hhofmann@physik.tu-muenchen.de}
}
\date{May 27, 1999}
\maketitle

\vspace{2cm}
\begin{abstract}
The linear response approach to nuclear transport has been
extended to pair correlations. The latter are treated within a
mean field approximation to a pairing interaction with constant
matrix elements $G$.  The constraint of particle number
conservation has been accounted for on a time dependent average,
which leads to modified  response functions, both in the pairing
degree of freedom as well as in the shape variable. The former is
expressed by the gap parameter $\Delta$ and the latter by a $Q$
which specifies the elongation of a fissioning nucleus. The
tensors for friction and inertia corresponding to these two
collective coordinates are computed along the fission path of
$^{224}Th$ for temperatures around $T=1 \mev$ and less. It is seen
that dissipation decreases with decreasing temperature and
increasing pairing gap and falls well below the values of common
"macroscopic models". Both friction and inertia show a sensible
dependence on the configurations of the mean field caused both by
shell effects as well as by avoided crossings of single-particle
levels.
\end{abstract}
\bigskip
\centerline{PACS numbers: -05.60.+w, 21.60.Cs, 21.60.Ev, 24.10Pa,
24.75+i }
\vfill

\newpage

\tableofcontents

\newpage

\section{Introduction}
\label{sec: intro}
Pair correlations are vital for understanding many elementary
features of nuclear physics at zero or small thermal excitations.
In particular, it is known that they have great impact on
collective properties, see e.g. \cite{bohrmo2} or \cite{rischu}.
For the case of zero temperature this has been demonstrated in
various ways. Here it may suffice to mention a few examples
related to nuclear fission. Since the early 70'ties it is known
that pairing will modify greatly the effective inertia
\cite{pauled}, which in turn may change the penetrability through
the barrier by orders of magnitude. In addition, the pair degree
of freedom in itself may portray important features, like pair
vibrations etc.\cite{bohrmo2}. A convenient way of handling such
properties is by introducing the gap parameter as an additional,
independent collective degree of freedom, which first has been
utilized in \cite{moretto}. Later this parameter has been
introduced also to the generator coordinate method \cite{stpipo},
where the influence of pairing vibrations on the spontaneous
fission probability has been studied.

It is of great interest to account for pair correlations also in
the description of typical transport problems of dissipative
systems. On general grounds it is to be expected that pairing will
greatly diminish nuclear dissipation. Indeed, there exist early
formulations within linear response theory where such effects have
been studied, see \cite{sijenlefre}-\cite{jenlefre}. The present
approach is meant as a generalization of these in a four fold
sense: (i) We will account for particle number conservation on a
time dependent average by introducing and exploiting modified
response functions. (ii) We will calculate response tensors (and
thus tensors for transport coefficients) for a pairing mode plus a
shape degree of freedom. (iii) This will be done along a complete
fission path parameterized in terms of Cassini ovaloids, thus
improving the results of \cite{ivhopaya} for pair correlations.
(iv) We will also address temperatures below $1\mev$, for which
pair correlations become even more important.

The paper is organized as follows: In Sec.~\ref{sec: LRT} we will
quote basic elements of linear response theory. For a more
detailed exposition of the theory we refer to the review article
\cite{hofrep}. Sec.\ref{sec: inclupair} contains the derivation of
the mean-field Hamiltonian and the response functions with pairing
included.  The Strutinsky renormalization will be applied to the
calculation of the free energy in Sec.~\ref{subsubsec: strut}.  In
Sec.\ref{subsec: gamma} the evaluation of the collisional width
for quasi-particles will be explained . In Sect.\ref{sec:
spurious} the modification of the response function will be given
which accounts for the average particle number conservation in the
dynamical process.  Finally, in Sec.\ref{sec: results} numerical
results for response functions and transport coefficients are
shown, the latter being compared in Sec.\ref{subsec: tdeptran}
with those of other approaches, with special emphasis put on the
temperature dependence of dissipation. Section \ref{sec: summ}
contains a short summary.
\section{ Linear response theory for collective motion}
\label{sec: LRT}
In this section we give a brief review of a description of collective
motion within the locally harmonic approximation (LHA). It is based on
the common hypothesis (see e.g. \cite{bohrmo2} and \cite{sije})
of the existence of a set of collective c-number variables $Q_{\mu}$
which in parametric way portray average dynamics of the nucleus as a
whole. Fluctuations in these quantities shall be neglected; how the
latter may be included can be found in \cite{hofrep}. The coupling of
these collective variables to the nucleonic degrees of freedom is
traced back to the deformed shell model potential which shows up in the
(many body) Hamiltonian $\hat H_{sm}$ for independent particle motion,
to be generalized later to the quasi-particle picture when pairing
correlation are to be included.

As is well known, the total energy of the system may be obtained from
$\hat H_{sm}$ after applying Strutinsky's renormalization procedure.
For later purpose it will be convenient to account for this fact by
adding to the $\hat H_{sm}$ a c-number term $\overline{{\cal E}}_{pot}$
to get
\bel{rmfconst}
\hat H_{rmf}(\hat x_i,\hat p_i,Q_{\mu}) =
\hat H_{sm}(\hat x_i,\hat p_i,Q_{\mu}) - \overline{{\cal E}}_{pot}~,
\end{equation}
which in some sense may be called the Hamiltonian of a
"renormalized mean field". At zero intrinsic excitations, the
additional term can be seen to be given by the negative value of
the average potential energy, see \cite{bohrmo2} and c.f.
\cite{sije}. Such a construction warrants that the total energy of
the system can be expressed by the mean value  $\bra \hat
H_{rmf}(\hat x_i,\hat p_i,Q_{\mu})\ket$. It may easily be extended
to finite temperature and the correction term may simply be found
by {\it requiring} that the energy obtained from the Strutinsky
procedure (at finite $T$) be identical to the one calculated from
$\hat H_{rmf}$, see \cite{hofrep}.

As mentioned earlier, we will be concerned with situations of
finite thermal excitations. In this case it must be expected that
the {\it pure} independent particle model breaks down even close
to the Fermi surface, in the sense that the particles (or
quasi-particles) experience "collisions".  Formally, the latter
may be represented by adding to the $\hat H_{rmf}$ a {\it residual
two-body interaction} which leads to the Hamiltonian \bel{hamilt}
\hat H(\hat x_i,\hat p_i, Q_{\mu})= \hat H_{rmf}(\hat x_i,\hat
p_i,Q_{\mu}) + \hat V^{(2)}_{res}(\hat x_i,\hat p_i)
\end{equation}
However, it is too cumbersome to work with such a form in genuine
sense. For this reason we are going to approximate the effects of the
$\hat V^{(2)}_{res}$ by dressing the single particle energies with
self-energies having both real and imaginary parts. A description of
details of this procedure will be deferred to a later section. We may
note here, that this approximation shall be done in such a way
that these self-energies are insensitive to changes in the
collective coordinates $Q_\mu$, for which reason the $Q_\mu$ have been
left out in the arguments of $\hat V^{(2)}_{res}$.

This latter approximation goes along with the conjecture that the
{\it "generators" of collective motion are given by the following one-body
operators}
\bel{generator}
\hat F_{\mu}({\hat x}_i,{\hat p}_i,Q_{\mu})\equiv
{\partial \hat H({\hat x}_i,{\hat p}_i,Q_{\mu})\over\partial Q_{\mu}}
\equiv  {\partial \hat H_{rmf}({\hat x}_i,{\hat p}_i,Q_{\mu})\over\partial
Q_{\mu}}
\end{equation}
These generators make up the coupling between collective and
intrinsic degrees of freedom. Within the LHA this may be seen as
follows, (c.f. \cite{hofrep}). Suppose the actual $Q_\mu$ is close
to some $Q_{\mu}^0$. One may then expand the effective Hamiltonian
to second order as \bel{hamilapp} \hat H(Q_{\mu}(t))=\hat
H(Q_{\mu}^0)  +
     \sum_{\mu}(Q_\mu(t) - Q_\mu^0)\h{F_{\mu}} +
     {1\over 2}\sum_{\mu\nu}(Q_\mu(t) - Q_\mu^0)(Q_\nu(t) - Q_\nu^0)
\left<{\partial^2\hat H\over \partial Q_{\mu}^0\partial Q_{\nu}^0}
    \right>_{Q_{\mu}^0,T^0}^{\rm qs}
\end{equation}
to describe general features in the neighborhood of the $Q_\mu^0$.
Evidently, such a procedure is possible for all static quantities.
In the dynamic case one needs to require that the $Q_\mu(t)$ does not
move away from $Q_\mu^0$  within some time $\delta t$ which is larger
than the typical time for the nucleonic degrees of freedom, say the
one which describes their relaxation to local equilibrium.

The generator $\h{F_{\mu}}$ appearing in \req{hamilapp} is the one
of \req{generator} but calculated at $Q_{\mu}^0$. It represents
the nucleonic part of the coupling between nucleonic and
collective degrees of freedom which is linear in the latter. As
implied by \req{hamilapp}, within the LHA this is the {\it only}
coupling term left. In the second order term the nucleonic part
appears only as a static average of the corresponding operator.
This average is to be built with the density operator
$\equiop(Q_{\mu}^0)$ which in the quasi-static picture is defined
by the Hamiltonian at $Q_{\mu}^0$, namely $\hat H(Q_{\mu}^0)$. The
$\equiop(Q_{\mu}^0)$ is meant to represent thermal equilibrium at
$Q_{\mu}^0$ with the excitation being parameterized by temperature
or by entropy. The simplest possibility is offered by using the
canonical ensemble, or more generally, the {\it grand canonical
distribution}.  For any (static) $Q_\mu$ the latter is defined as
\bel{rhoqs} \rho_{\rm qs}(Q_{\mu},T,\mu)={1\over Z^\prime}
\exp(-\hat H^\prime(\hat x_i,\hat p_i,Q_{\mu},\mu)/T), \quad
Z^\prime= {\rm tr}\,(\exp(-\hat H^\prime(\hat x_i,\hat
p_i,Q_{\mu},\mu)/T))
\end{equation}
with
\bel{hgran}
\hat H^\prime(\hat x_i,\hat p_i,Q_{\mu},\mu) =
\hat H(\hat x_i,\hat p_i,Q_{\mu}) - \mu \hat N
\end{equation}
and $\mu,~\hat N$ being the chemical potential and the particle
number operator, respectively.

For the remaining part of this section we are going to restrict
ourselves to a fixed $N$ and discard complications from a possible
variation of particle number. Later we are going to study the
general case for which $\mu$ is considered just another
"collective variable".

Before we address time dependent forces let us quote a general relation
for the generalized static ones along some given direction $Q_\mu$.
With $E(Q_\mu,S)$ being the internal energy at fixed entropy $S$ and
the ${\cal F}(Q_\mu,T)$ the free energy at given temperature
this relation reads
\bel{firstvaren}
\left<{\partial\hat H\over\partial Q_{\nu}} \right>^{\rm qs}
 = \left({\partial{\cal F}(Q_\nu,T)\over \partial Q_{\mu}}\right)
  _{T,Q_{\nu\neq \mu}}
  = \left({\partial E(Q_\mu,S)\over \partial Q_{\nu}}\right)
 _{S,Q_{\nu\neq \mu}}
\end{equation}
It is valid at any $Q_\mu$ both at or away from global minimum where
these derivatives vanish (see \cite{kidhofiva} and \cite{hofrep}, again).

The equations of average motion for $Q_{\mu}(t)$ may
be constructed looking at energy conservation. For an arbitrary path
through the multi-dimensional collective space the change of the total
energy can be expressed as
\bel{enercons}
0 = {d \over dt} E_{\rm tot} = \sum_{\mu}{\dot Q_{\mu}} \left<{\partial
\ham{Q} \over \partial Q_{\mu}}\right>_{t}
\end{equation}
It must vanish if we picture the nucleus as a whole as an isolated
system. Requiring this condition to hold true for {\it any path}
through the collective landscape one gets the set of equations
\bel{eqofmotim}
0= \left<{\partial \ham{Q} \over \partial Q_{\mu}}\right>_{t} =
\left< \hat F_{\mu}\right>_{t}\; +
\sum_{\nu}(Q_\nu(t) - Q_\nu^0)
\left<{\partial^2\hat H\over \partial Q_{\mu}^0\partial Q_{\nu}^0}
    \right>_{Q_{\mu}^0,T^0}^{\rm qs}
\end{equation}
To put them into convenient form one first needs to express the
average $ \left< \hat F_{\mu}\right>_{t}$ as a functional of
$Q_{\mu}(t)$ as well as a function of possible changes in either
temperature or entropy. The form of the coupling term as given in
\req{hamilapp} invites one to apply linear response theory to get the
actual time dependent part. Considering
$\sum_{\mu}\hat F_{\mu}(Q_\mu(t) - Q_\mu^0)$ as a time dependent
perturbation one may express the deviation of the $\bra \hat F_{\mu}\ket_t$
from their quasi-static values as (see \cite{hofrep})
\bel{delft}
\delta \bra \hat F_{\mu}\ket_t =
-\sum_{\nu}\int_{-\infty}^{\infty}
\tilde\chi_{\mu\nu}(t-s)(Q_\nu(s) - Q_\nu^0)ds
\end{equation}
where $\rft_{\mu\nu}$ is the causal response function
\bel{defrest}
\rft_{\mu\nu} (t-s)= \Theta(t-s) {i\over \hbar}
             {\rm tr}\,\left(\hat{\rho}_{\rm qs}(Q_{0},T_{0})
             [\hat F_{\mu}^{I}(t),\hat F_{\nu}^{I}(s)] \right)
     \equiv 2 i \Theta(t-s) \drft_{\mu\nu}(t-s)
\end{equation}
Here, the time dependence of $\hat F_{\mu}^{I}(t)$ (in interaction
representation, if one wishes) is determined by the
$\hat H(Q_{\mu}^0)$.
The Fourier transform of \req{delft} reads
\bel{delfcoll}
\delta \bra \hat F_\mu\ket_{\om}=
-\sum_{\nu}\chi_{\mu\nu}\delta Q_{\nu}(\om)
\qquad {\rm with} \qquad
\delta Q_{\nu}(\om)=Q_\nu(\om) - Q_\nu^0\delta (\om)
\end{equation}
with $\delta Q_{\nu}(\om)$ being the Fourier transform of $(Q_\nu(s) -
Q_\nu^0)$.

Piecing together all relevant parts, in Fourier representation and
exploiting vector notation the set of equations of motion may
finally be written in the following compact form \bel{manyself}
\delta\bra{\bf \hat F}\ket_{\om}= {\bf k^{-1}}{\bf q}(\om)
     \qquad {\rm with} \qquad
{\bf q}={\bf Q} - {\bf Q}_m
\end{equation}
and where the  $Q_\mu^m$ defines the center of the local
oscillators (see \cite{hofrep} where all details have been worked
out carefully for the one-dimensional case). In \req{manyself} a
tensor $k_{\mu \nu}$ appears whose elements often are interpreted
as coupling constants, see below. At zero excitation it would
simply be given by the negative value of the average of the second
derivatives of the Hamiltonian \cite{sije}. At finite excitations
additional terms show up which involve static susceptibilities.
Which ones will appear depends on the situation one chooses. In
\cite{hofrep} it has been argued in favor of working at constant
entropy. It is much simpler, though, to assume that temperature
does not change within the time lapse $\delta t$ for which the
equations of collective motion are derived. For the sake of
simplicity we will adhere to the latter case, knowing that in this
respect we are along the lines of most of the papers in this
field.  For such a situation the coupling tensor, which often is
expressed by its inverse $\mbox{{\boldmath
$\kappa$}}\equiv\mbox{{\boldmath $k$}}^{-1}$, is given by (c.f.
\cite{hofrep}) \bel{coupconst} -\left(k^{-1}\right)_{\mu\nu}\equiv
-\kappa_{\mu\nu}=
 \left<{\partial^2\hat H\over\partial Q_{\mu}\partial Q_{\nu}}
 \right>^{\rm qs} + \chi_{\mu\nu}(0)-\chi_{\mu\nu}^{\rm T}=
   {\partial^2{\cal F}(Q,T)\over \partial Q_{\mu}\partial Q_{\nu} }
   +\chi_{\mu\nu}(0)
\end{equation}
Here, $\chi_{\mu\nu}(0)$ is the static response and
\bel{isothesus}
\chi_{\mu\nu}^{\rm T}=\left.
-{\partial\bra\hat F_{\mu}\ket^{\rm qs} \over
\partial Q_{\nu}}\right\arrowvert_T
\end{equation}
the isothermal susceptibility defined as the change of the
quasi-static expectation value $\bra\hat F_{\mu}\ket^{\rm qs}$
with $Q_{\nu}$ at given $T$.  It may be noted that at lower
temperature the difference $\chi_{\mu\nu}(0)-\chi_{\mu\nu}^{\rm
T}$ --- known to vanish exactly at $T=0$ --- may often be
neglected as compared to the first term on the right of
\req{coupconst}. Indeed, such a case will be encountered below.

Finally, we aim at equations of motion for the $Q_\mu$ of a structure
similar to that of Newton's equation with a dissipative force.
To be able identifying uniquely all forces, as well as the
corresponding coefficients for inertia, dissipation and local
stiffness, it is useful to introduce a coupling $U^{ext}(t)$ to
some "external" fields $f_{\mu}^{ext}(t)$ which finally will
show up as in-homogeneous terms. A convenient form is given by
\bel{fext}
U^{ext}(t)=
\sum_{\mu}\hat F_{\mu}f_{\mu}^{ext}(t)
\end{equation}
Defining a response tensor $\mbox{{\boldmath $\chi$}}_{\rm coll}(\om)$
by
\bel{delfcollom}
\delta \bra \hat F_{\mu}\ket_{\om}=
-\sum_{\nu}\chi_{\mu\nu}^{\rm coll}(\om) f_{\nu}^{ext}(\om)
\end{equation}
it can be shown (see \cite{sahoya} and \cite{hofrep}, for instance, or
\cite{sije} for the undamped case at $T=0$) to attain
the form
\bel{collresp}
\mbox{{\boldmath $\chi$}}_{\rm coll}(\om)=
\mbox{{\boldmath $\kappa$}}(\mbox{{\boldmath $\kappa$}}+
\mbox{{\boldmath $\chi$}}(\om))^{-1}\mbox{{\boldmath $\chi$}}(\om)
\end{equation}
Its pole structure is determined by the collective excitations of the
system, whence the name "collective" response tensor comes from.

To finally be able introducing the transport coefficients for
average collective motion, as expressed in terms of the
$q_\mu(t)$, one first rewrites \req{delfcollom} in the form
\bel{quasieom} [\mbox{{\boldmath $\kappa$}}\mbox{{\boldmath
${\chi}$}}_{\rm coll} \,^{-1}(\om)\mbox{{\boldmath
$\kappa$}}]\mbox{{\boldmath $q$}}(\om) =-\mbox{{\boldmath
$\kappa$}}\mbox{{\boldmath $f$}}^{ext}(\om) \equiv
-\mbox{{\boldmath $q$}}^{ext}(\om)
\end{equation}
Here, \req{manyself}  has been used and the external fields $q^{ext}$
for the $q_\mu$-modes are defined through the last equation on the very
right. The set \req{quasieom} turns into the conventional form of
Newtonian dynamics
\bel{classeom}
\sum_{\nu}(M_{\mu\nu}\ddot q_{\nu}(t)+
\gamma_{\mu\nu}\dot q_{\nu}(t)+C_{\mu\nu}q_{\mu}(t))=-q^{ext}_{\mu}(t)
\end{equation}
if one may approximate the quantity
$\mbox{{\boldmath $\kappa$}}\mbox{{\boldmath $\chi$}}_{\rm coll}
\,^{-1}(\om)\mbox{{\boldmath $\kappa$}}$ by a second order
polynomial in frequency
\bel{lorfit}
(\mbox{{\boldmath $\kappa$}}\mbox{{\boldmath $\chi$}}_{\rm coll}
\,^{-1}(\om)\mbox{{\boldmath $\kappa$}})_{\mu\nu}\quad
\Longrightarrow\quad -M_{\mu\nu}\om^2-i\gamma_{\mu\nu}\om+C_{\mu\nu}
\end{equation}
Evidently, the coefficients $M_{\mu\nu}$, $\gamma_{\mu\nu}$ and
$C_{\mu\nu}$ stand for the elements of the tensors for mass, friction
and stiffness.

Without any doubt, the transition \req{lorfit} is crucial
to the question as to which extent Markovian type of equations
may be justified for the very complex situation of nuclear collective
motion. In the past, two version have been worked out for this
transition (see \cite{hofrep} for a review and a guide to original
work). On the one hand, one may study the collective strength
distribution as given by the dissipative part of the collective
response. Whenever this distribution has prominent peaks (for the
multi-dimensional case see \cite{sahoya}) they may be fitted by
Lorentzians, which in turn may be interpreted as the oscillator
response function associated to the form \req{classeom} of the
classical equations of motion. Another possibility is offered by
simply expanding the functions given on the left hand side of
\req{lorfit} to second order in $\om$ around $\om=0$. In the
one-dimensional case one gets, see
\cite{ivhopaya}),
\begin{equation}
\label{colzerstif}
C \approx
    {1\over k^2 \chi_{\rm coll}(\omega)}\Bigr\arrowvert_{\omega=0}=
    {{\chi (0)+C(0)}\over{\chi(0)}}C(0)
\end{equation}
\begin{equation}
\label{colzerfric}
\gamma \approx {1\over k^2}
    {\partial (\chi_{\rm coll}(\omega))^{-1} \over \partial \omega}
    \Bigr\arrowvert_{\omega=0}=
    {{(\chi (0)+C(0))^2}\over{\chi^2(0)}}\gamma(0)
\end{equation}
and
\begin{equation}
\label{colzermass}
M \approx {1\over 2k^2}{\partial^2 (\chi_{\rm coll}(\omega))^{-1}
    \over \partial \omega^2} \Bigr\arrowvert_{\omega=0}= {{(\chi
(0)+C(0))^2}
    \over {\chi^2(0)}} \left(M(0)+{\gamma^2(0)\over \chi (0)}\right)
\end{equation}
The friction $\gamma(0)$ and mass parameters $M(0)$ are expressed
in terms of first and second derivatives of the response function
at $\om=0$. Their modification to the multi dimensional case are
of the forms
 \bel{zerofrilim} \gamma_{\mu\nu}(0)=
-i{{\partial\hat\chi_{\mu\nu}(\omega)}\over{\partial \omega}}
\Bigr\arrowvert_{\omega = 0}={{\partial
\hat\chi_{\mu\nu}^{\dpr}(\omega)}
    \over{\partial \omega}}    \Bigr\arrowvert_{\omega = 0}
\end{equation}
and \bel{zeromass} M_{\mu\nu}(0)={1\over
2}{{\partial^2\hat\chi_{\mu\nu}(\omega)} \over{\partial
\omega^2}}\Bigr\arrowvert_{\omega = 0}= {1\over
2}{{\partial^2\hat\chi_{\mu\nu}^{\pr}(\omega)} \over{\partial
\omega^2}}\Bigr\arrowvert_{\omega = 0}.
\end{equation}
For obvious reasons, expressions \req{zerofrilim},\req{zeromass}
may be referred to as the "zero frequency limit".

\section{The inclusion of pairing degrees of freedom}
\label{sec: inclupair}
In this section we are going to address pair correlations. This shall be
done in two steps. At first we concentrate on the pairing mode alone to
subsequently address the general case of pairing plus a shape degree of
freedom.
\subsection{The conventional pairing model}
\label{subsec: convpair}
The generator for the pairing mode may be defined as the following
pairing field operator $\hat P^\dagger$
\bel{pair}
\hat P^\dagger = \sum_k\akp\akbp.
\end{equation}
The $\akp$ and $a_k$ are the creation and annihilation operators
for the normal vacuum. The summation over single-particle states
$k$ is meant not to include the time reversed ones, which are
denoted by ${\bar k}$.  Commonly one does not start with the
associated mean field Hamiltonian, as we did in the first
section, but with a separable two body interaction instead,
which may be constructed from the $\hat P$ and  $\hat
P^\dagger$.  The corresponding many body Hamiltonian  may thus
be written as
\bel{hamil}
\tbhp={\hat H}_{sm} - G\hat P^\dagger\hat P
\end{equation}
where the coefficient $G$ of the two body interaction, the coupling
constant, stands for a constant pairing matrix.
The first part ${\hat H}_{sm}$ is meant to represent
motion of {\it independent particles}, and hence may be written as
\bel{hipm}
{\hat H}_{sm}=\sum_k \epsk (\akp a_k + \akbp \akb)
\end{equation}
with the $\epsk$ being the single-particle energies.
How this ${\hat H}_{sm}$ relates to the shape degrees of freedom
shall be of no concern for the moment. As indicated earlier this
question will be addressed later.

To establish relations to the discussion in the previous section let
us introduce the mean field approximation to \req{hamil} by writing
\bel{hamilmod}
\tbhp={\hat H}_{sm} - \Delta (\hat P^\dagger + \hat P) +{\Delta^2\over G}
+ {\hat H}_{res}
\end{equation}
with
\bel{delta}
\Delta\equiv G\bra\hat P\ket=G\bra\hat P^\dagger\ket
\end{equation}
and
\bel{hone}
\hat H_{res} =
-G\sum_{kj}(\akp\akbp-\bra\akp\akbp\ket)(\ajb a_j-\bra\ajb a_j\ket)
\end{equation}
Nothing else has been done but rewriting \req{hamil} with the help
of the c-number variable $\Delta$. Actually at this stage it may
be unclear yet whether or not the averages of $P$ and $P^\dagger$
are identical. In the worst case we might simply introduce a
complex $\Delta$. However, as we shall see a real one will do.

Evidently, the mean field part of \req{hamilmod} may be identified as
\bel{hamilmf}
\tbhp_{rmf}={\hat H}_{sm} - \Delta (\hat P^\dagger + \hat P)
+{\Delta^2\over G}
\end{equation}
Actually, in spirit of the last section the word "renormalized" is
appropriate here; it is justified because of the c-number term on
the very right. Indeed, this term is necessary in order to have
the expectation value of $\tbhp_{rmf}$ represent the total energy,
see e.g. sect.3.1.5 of \cite{hofrep}, in particular eq.(3.1.45).
For the present model this c-number term is nothing else but the
$\overline{{\cal E}}_{pot}$ of \req{rmfconst}. In summary, the
mean field approximation to \req{hamil} simply implies to neglect
the residual interaction $H_{res}$; see also \cite{jenlefre} or
\cite{sanyam}.

One might be inclined to associate the $\hat H_{res}$ of
\req{hone} with the $\hat V^{(2)}_{res}$ introduced in
\req{hamilt}. This is not the appropriate connection, however, as
the $\hat V^{(2)}_{res}$ is meant to simulate collisional damping.
There can be little doubt that it would be asking too much if one
were to deduce the latter from a separable interaction of the type
given in \req{hamil}. Indeed, later on we are going to take into
account this latter type of residual interaction, albeit in
special form, namely through self-energies, which in some sense
may be considered a generalized mean field approximation.

The $\Delta$ introduced in \req{hamilmod} will henceforth be
considered the collective variable for the pairing degree of
freedom and the associated mean field Hamiltonian will be the one
of \req{hamilmf}. The alert reader may have noticed some
similarity of \req{delta} with \req{manyself}, indicating a
possible connection of the $G$ to the coupling constant $k$
introduced in the previous section; this latter relation will be
worked out in more detail below. Notice, however, an essential
difference between \req{manyself} and \req{delta}. Whereas the
former case refers to a time dependent situation in which the
expectation values have a definite meaning. Those appearing in
\req{delta}, on the contrary, are not yet fully specified. Indeed,
the Hamiltonian given in \req{hamilmf} is of practical use only if
the $\Delta$ may at first be treated as a {\it free parameter ---
as it ought to be for a collective variable --- discarding this
subsidiary condition \req{delta}}. It is only after the $\Delta$
has been assigned a special value (which for the static case will
be that of global equilibrium), or a well defined function of time
(as in the dynamic case discussed above for the general situation,
with its specification to the case of pairing to come below) that
these expectation values will be defined properly.

\subsubsection{Transformation to quasi-particles}
\label{subsec: traquap}
In the following we are going to diagonalize the Hamiltonian
$\tbhp_{rmf}$ by transforming to quasi-particles in strict
sense.  In doing so we will not make use of the
subsidiary condition \req{delta}. The relations to be
discussed now will hold true in the more general sense, not just
at equilibrium which we will look at below. Actually it is at this
place that we exploit the fact of $\Delta$ being a free
parameter.

When accounting for pair correlations it is difficult to conserve
particle number. For this reason we will refer to the {\it grand
canonical ensemble} introduced in \req{rhoqs} and \req{hgran}.
This means to replace the $\tbhp_{rmf}$ by \bel{hpr} \tbhpP_{rmf}
=\tbhp_{rmf}-\mu \hat N
\end{equation}
fixing the Lagrange multiplier $\mu$ by requiring particle number
conservation on average, viz
\bel{partop} \langle \hat N \rangle =
N \qquad {\rm with} \qquad \hat N=\sum_k (\akp a_k + \akbp \akb)
\end{equation}

To diagonalize the Hamiltonian $\tbhpP_{rmf}$ let us perform
first the Bogolyubov transformation
\bel{aas}
a_k=u_k \alk + v_k \alkbp  \qquad \akb=u_k \alkb - v_k \alkp
\end{equation}
to new operators $\alk, \alkp\ldots$ which themselves are required
to satisfy the anti-commutation rules of Fermions. As one knows,
this implies the constraints
\bel{uksqvksq} u_k^2 +v_k^2 = 1.
\end{equation}
Next we write
\bel{hprqp}
\tbhpP_{rmf}=U^\prime_0 + H^\prime_{11} + H^\prime_{20}
\end{equation}
where
\begin{eqnarray}\label{u0h11h20}
U^\prime_0=\sum_k 2v_k^2 (\epsk -\mu) -2\Delta\sum_k u_k v_k+{\Delta^2\over G}
\,\,, \nonumber\\
H^\prime_{11}=\sum_k[(\epsk-\mu)(u_k^2-v_k^2)+2\Delta u_k v_k)]
(\alkp\alk+\alkbp\alkb)\,\,,\nonumber\\
H^\prime_{20}=\sum_k[(\epsk-\mu)2u_k v_k-\Delta(u_k^2-v_k^2)]
(\alkp\alkbp+\alkb\alk)
\end{eqnarray}
following common notation. To get from \req{hprqp}-\req{u0h11h20}
the independent quasi-particle Hamiltonian we must chose the
coefficients $u_k$ and $v_k$ of the transformation \req{aas} to
make the last term vanish: $H^\prime_{20}=0$.  As it is seen from
\req{u0h11h20}, this condition has the form \bel{h20eq0} (\epsk
-\mu)2u_k v_k-\Delta(u_k^2-v_k^2)=0
\end{equation}
This equation, together with normalization condition
\req{uksqvksq}, can be solved as \bel{ukvk} u_k^2={1\over
2}\left(1+{\epsk -\mu\over{\sqrt{(\epsk -\mu)^2+\Delta^2}}}
\right),\qquad v_k^2={1\over 2}\left(1-{\epsk
-\mu\over{\sqrt{(\epsk -\mu)^2+\Delta^2}}} \right)
\end{equation}
Finally, the $\tbhpP_{rmf}$ becomes
\bel{hiqp}
\tbhpP_{rmf}=U_0^\prime+\sum_k E_k (\alkp\alk+\alkbp\alkb)
\end{equation}
with
\bel{Ek}
E_k = \sqrt{(\epsk -\mu)^2+\Delta^2}.
\end{equation}
\subsubsection{Thermodynamics of quasi particles}
\label{subsec: thermo}
Having the quasi-particle Hamiltonian $\tbhpP_{rmf}$ of \req{hiqp}
at our disposal we may now look at the density operator for the
{\it quasi-static} equilibrium for the pairing degrees of freedom.
It is a special case of \req{rhoqs} it reads \bel{quastair}
\rho_{\rm qs}(\Delta,T)={1\over Z^\prime} \exp(-\tbhpP_{rmf}/T)
\qquad\qquad Z^\prime= {\rm tr}\,\exp(-\tbhpP_{rmf}/T)
\end{equation}
(Please observe that the $\Delta$ still is to be considered a free
parameter). For a Hamiltonian like the one of \req{hiqp}
representing independent motion of quasi-particles it is not
difficult to calculate various quantities of interest.  For the
grand potential one gets \bel{grand} {\Om}=-T\log Z^\prime =\sum_k
(\epsk -\mu-E_k) -2T\sum_k\log(1+\exp(-E_k/T))+{\Delta^2\over
G}\,,
\end{equation}
and for the particle number
\bel{partnum}
N=-{\partial \Om\over\partial\mu}= 2 \sum_k n_k
\end{equation}
where
\bel{nk}
 n_k= \langle\akp a_k\rangle^{\rm qs} = v_k^2 +(u_k^2-v_k^2)n_k^T =
{1\over 2}\left(1-{\epsk -\mu\over E_k}\tanh{E_k\over 2T}\right),
\end{equation}
with
\bel{thermocc}
n_k^T=\langle\alkp\alk\rangle^{\rm qs}=(1+\exp(E_k/T))^{-1}
\end{equation}
being the probability that the state $k$ is excited.
Let ${\cal P}$ be the average value of the pairing field
operator for the ensemble \req{quastair};
one finds
\bel{calp}
{\cal P}\equiv\langle \hat P\rangle^{\rm qs}\equiv
\bra \hat P^\dagger\ket^{\rm qs}=\sum_k\phi_k
\end{equation}
with
\bel{phik}
\phi_k=\bra \akp\akbp\ket^{\rm qs}=u_k v_k(1-2n_k^T)={\Delta\over
2E_k}(1-2n_k^T).
\end{equation}
The internal energy may be written as
\bel{energy}
E=\bra \tbhpP_{rmf}\ket^{\rm qs} +\mu N \equiv \bra \tbhp_{rmf}\ket^{\rm
qs}
=2\sum_k \epsk n_k -2\Delta{\cal P}+{\Delta^2\over G}
\end{equation}
and for the free energy one gets
\bel{freeen}
{\cal F}=\Om +\mu N =
2\sum_k \epsk n_k -2\Delta{\cal P}+{\Delta^2\over G} -TS
\end{equation}
with the entropy $S$ being given by
\bel{sk}
S = 2\sum_k s_k
   \qquad {\rm with} \qquad
s_k =[n_k^T \log n_k^T + (1-n_k^T)\log(1-n_k^T)]
\end{equation}

At various places we need the derivatives with respect to $\Delta$
rather than the energies themselves. In principle, they could be
obtained from \req{energy} and \req{freeen}. It is much easier,
however, to exploit eq.\req{firstvaren}. For the pairing degree of
freedom, and for the present purpose, the appropriate Hamiltonian
to be used there is the $\tbhp_{rmf}$ of \req{hamilmf}. Obviously,
its derivative is given by \bel{deripamf} {\partial
\tbhp_{rmf}(\Delta)\over\partial \Delta}= -(\hat P+\hat
P^\dagger)+{2\Delta\over G}
\end{equation}
Thus \req{firstvaren} leads to
\bel{varendel}
 \left({\partial{\cal F}(Q,\Delta,T)\over \partial \Delta}\right)
  _{T,Q} =
 -2 {\cal P} +{2\Delta\over G}
  = \left({\partial E(Q,\Delta,S)\over \partial \Delta}\right)
_{S,Q}
\end{equation}
It is reassuring to see that the same result may be obtained starting
from the expression for the free energy given in \req{freeen}, indeed.

Obviously, this force vanishes at
\bel{globequi}
{\cal P}^{\rm eq} \equiv {\cal P}(\Deleq) ={\Deleq\over G}
\end{equation}
thus defining the value $\Deleq$ at which either the internal
energy at fixed entropy or the free energy at fixed temperature
become minimal. Notice that \req{globequi} is an {\it implicit
equation} for $\Deleq$ with the functional form of ${\cal
P}(\Delta)$ being given by \req{calp}, together with \req{phik} and
the expressions given before for $E_k$ as well as for the thermal
occupation number: Eq.\req{globequi} is nothing else but the {\it
gap equation} at finite temperature, which can easily be brought
to the conventional form
\bel{gengapeq} {2\over G}  = \sum_k {1-
2n_k^T\over E_k} = \sum_k {\tanh(E_k/(2T))\over E_k}
\end{equation}

As one may recognize, the relation \req{globequi} is of the form
given in \req{delta}, which in this context has obtained a
specific meaning. For a static situation, the averages appearing
in \req{delta} are those of thermodynamic equilibrium, in which
case the subsidiary condition \req{delta} is valid for $\Delta$
replaced by $\Deleq$. As this special form is obtained after
minimizing the total energy, one may understand why sometimes it
is referred to as the "self-consistency condition" (for the
separable interaction we started with).

\subsubsection{Remarks on the coupling constant}
\label{subsec: remcoco}
In \req{coupconst} a coupling constant $k$ has been introduced which
depends on the collective variables as well as on excitation. Let us see how
this $k$ relates to the $G$ of \req{hamil}. A direct application of
\req{coupconst}  leads to
\bel{kappadd}
-{1\over k_\Delta}=
\left<{\partial^2 \tbhp_{rmf}(\Delta)\over\partial \Delta^2}\right>^{\rm
qs} +(\chi_{PP}(\om=0)-\chi^T_{PP}) =
{2\over G}+ (\chi_{PP}(\om=0)-\chi^T_{PP})
\end{equation}
For the second equation we have simply made use of \req{deripamf}.
The response function and the isothermal susceptibility appearing
there may be traced back to \req{defrest} and \req{isothesus} if
only the $\hat F_\mu$ is replaced by $-(\hat P +\hat P^\dagger)$.
Within the independent quasi-particle model their calculation is
straightforward. (The evaluation of the response in the more
general context will be given below, in particular in the
Appendix). As the susceptibility may be written as
$\chi_{PP}^{T}=2{\partial{\cal P} /\partial \Delta}$ it may be
obtained by differentiating \req{calp} with respect to $\Delta$
keeping temperature and particle number fixed. One gets
\bel{chittdd} \chi^T_{PP}=\widehat\chi_{PP}(0) - 2\sum_k{\Delta\over
E_k}{\partial n_k^T\over\partial \Delta}
\end{equation}
(The $\widehat\chi_{PP}(\om)$ is the response function calculated for
fixed particle number, for details see below).
Estimating numerically the difference $\chi^T_{PP} -\chi_{PP}(\om=0)$
from \req{chittdd} it is seen to be negligibly small as compared to the
leading term $2/G$. Even at $T=2\mev$ it makes up less than $10\%$.
Exploiting this fact we may conclude
\bel{paircoco}
-k_\Delta \approx G/2\; ,
     \qquad {\rm to~put} \qquad -k_\Delta \equiv G/2
\end{equation}
Evidently, for this comparison of $k_\Delta$ and $G$, there is no
need to worry about the different signs or the appearance of the
factor of $2$, which are simply due to the fact of using in
\req{hamil} a form commonly found in the literature. However, a
few words might be in order to explain why this relation has
turned out not to be an exact one. First of all, we may recall,
that at $T=0$ this would be so in strict sense. Secondly, it had
been mentioned already when discussing the appearance of the $k$
in \req{manyself}, that for the nucleus as an isolated system a
more appropriate form would be the one in which in \req{coupconst}
the isothermal susceptibility gets replaced by the adiabatic one
$\chi^{\rm ad}$ defined for fixed entropy. This difference,
$\chi^{\rm ad}-\chi(0)$, is known to vanish identically for
ergodic systems, see \cite{hofrep}, in which case $-k_\Delta$
equals $G/2$ exactly, indeed. Thirdly, it can be said for {\it
diabatic motion of independent quasi-particles} this difference
again would vanish identically. This observation follows from the
fact for such a model diabatic motion is given for {\it fixed
occupation numbers} in which case the $n_k^T$ do not vary with the
collective variables (like the $\Delta$ in the present case).

\subsection{The general mean field for pairing and shape variables}
\label{subsec: general}
After this detour into the peculiarities of the pairing model we
may now go back to the starting point in Sec.~\ref{sec: LRT} to
combine the collective variable $\Delta$ for pairing with those for
the shape. Actually, to keep the numerical effort on a manageable
level, we restricted the computations to just one shape variable,
the elongation parameter $Q=R_{12}/2R_0$ which describes the
distance between the left and right centers of mass. To have a
dimension-less quantity this distance is commonly divided by the
diameter of the sphere of identical volume, see \cite{ivhopaya}.

To construct the Hamiltonian $\hat H_{sm}(\hat x_i,\hat
p_i,Q_{\mu})$ of \req{rmfconst} for the set  $Q_{\mu}\equiv
Q,\Delta$, we simply need to choose for the $\hat H_{sm}$ of
\req{hamilmf} the Hamiltonian $\hat H_{sm}(\hat x_i,\hat p_i,Q)$
of the deformed shell model.  The quasi-static density operator
for the grand canonical ensemble is given by \req{rhoqs}.  The
generators of collective motion defined in \req{generator},
specialized now to $Q_{\mu}\equiv Q,\Delta$, turn out to be as
follows. The one for the pairing mode is given already by
\req{deripamf}. Let us write it as\bel{hatfd}
\hat F_{\Delta}=
{\partial \hat H_{rmf}(Q,\Delta)\over\partial \Delta}=
-(\hat P+\hat P^\dagger)+{2\Delta\over G}
\end{equation}
It explicitly contains the $c$-number correction $2\Delta/G$.
The equivalent form for the shape degree of freedom reads
\bel{hatfq}
\hat F_{Q}=
{\partial \hat H_{rmf}(Q,\Delta)\over\partial Q}=
{\partial \hat H_{sm}(Q,\Delta)\over\partial Q}\Bigr\vert_{Q=Q_0} -
 {\partial \overline{{\cal E}}_{pot} \over\partial
 Q}\Bigr\vert_{Q=Q_0}  \equiv \hat F -
 {\partial \overline{{\cal E}}_{pot} \over\partial
 Q}\Bigr\vert_{Q=Q_0}
\end{equation}
where the last term in the end would have to be calculated through the
Strutinsky renormalization, see below. It should be noted however,
that these corrections may be omitted when calculating the response
functions. They simply drop out of the commutators in \req{defrest}.

Next we turn to the coupling constants \req{coupconst}. Again,
following the discussion of \req{paircoco} the one
for the pairing degree is solely determined by $G$, in that we have
$-k_\Delta \equiv G/2\equiv -1/\kappa_\Delta$.
The one for the shape is
given by
\bel{appaqq}
-\kappa_{Q}=
{\partial^2{\cal F}(Q,\Delta,T)\over \partial Q^2}+\chi_{FF}(\om=0)
\end{equation}
This one will depend both on our collective variables $Q$ and $\Delta$
as well as on temperature $T$. In principle, there could be a
non-diagonal one. Following strictly the formulas given above we would find
\bel{appaqd}
-\kappa_{Q\Delta}=
{\partial^2{\cal F}(Q,\Delta,T)\over \partial Q\partial\Delta}
-\chi_{FP}(\om=0) = \chi^T_{FP}-\chi_{FP}(\om=0)
\end{equation}
The expression on the very right of \req{appaqd} follows from
\req{varendel} once its $Q$-dependence is taken into account. We
find it wiser, however, to simply put this quantity equal to zero,
essentially for two reasons, beyond that of simplicity: 1) Like
before for $\Delta\Delta$ case, the difference of the
susceptibilities $\chi_{FP}^T-\chi_{FP}(\om=0)$ can be expected to
be small at small $T$. 2) Again, we are faced with the fact that
the shape and the pairing degrees of freedom are not really based
on the same footing --- which by the way is a feature common to
almost all microscopic approaches. On the one hand, for the
pairing degree of freedom there is no mean field at our disposal
which might be considered as realistic as that for the shape
degrees of freedom (completed by the Strutinsky renormalization).
As a matter of fact, it was constructed from a separable two body
interaction. 3) Conversely, for the shape degrees of freedom it
would not make much sense to start from a similar schematic force.
We simply believe our approximating the mean field in this case to
be better.

\subsubsection{Strutinsky renormalization}
\label{subsubsec: strut}

Let us address the  calculation of the free energy ${\cal F}$,
needed for instance for the coupling constant $\kappa_{Q}$ given
in \req{appaqq}.  One might exploit the expression \req{freeen}
derived within independent quasi-particles model.  As is well
known, such an approximation holds true only in the neighborhood
of the Fermi level. But such deficiencies can easily be overcome
by applying the Strutinsky renormalization procedure \cite{strut,
brdajepastwo}. Thus we represent ${\cal F}$ as the sum of the
liquid drop part and the shell correction \bel{ftot} {\cal
F}={\cal F}_{LDM} + \delta {\cal F}
\end{equation}
Unfortunately, we can not use here the standard expression for the
shell correction to the {\it ground state energy} given in
\cite{brdajepastwo}. The derivation of this expression rely very
much on the gap equation which does not hold away from equilibrium
with respect to $\Delta$. For an arbitrary $\Delta$ we should
proceed in slightly different way. Namely, we may make use of the
advantage of Strutinsky's energy theorem according to which the
shell correction to the total energy of the system of interacting
particles can be obtained within a mean field approximation. In
our case it means that we can use the free energy derived not with
the Hamiltonian \req{hamil} but with independent quasi-particles
Hamiltonian \req{hiqp}. Namely, we will define the shell
correction to the free energy as \bel{deftaf} \delta{\cal F}={\cal
F}-\widetilde{\cal F}
\end{equation}
where ${\cal F}$ is given by \req{freeen} and $\widetilde{\cal F}$ is
its average.

The numerical computation of the average free energy
${\cal F}$ is done in two steps. Firstly, in eqs.\req{freeen} we
substitute summation over the single-particle states by
integration over the  single-particle energies with their density of
states being that of the independent particle model
\bel{sumint}
\sum_k \rightarrow \int_{-\infty}^{\infty}g_s(e)de\,\,\,\,{\rm where}\,\,\,\,
g_s(e)=\sum_k \delta(\epsk -e)
\end{equation}
In a second step we replace the density $g_s(e)$ by the smoothed quantity
\bel{gstogt}
g_s(e) \rightarrow \widetilde g(e)={1\over\gav}\int_{-\infty}^{\infty}
g_s(\ep)f_{av}\left({\ep-e\over\gav}\right)d\ep
={1\over\gav}\sum_k f_{av}\left({\epsk-e\over\gav}\right)
\end{equation}
where $f_{av}(x)$ is the smoothing function of the shell
correction method, see \cite{strut, brdajepastwo}.  In this way
the smoothed free energy becomes \bel{freeenav} \widetilde {\cal
F}=2\sum_k \epsk \widetilde n_k -
2\Delta\sum_k\widetilde\phi_k+{\Delta^2\over G} -2T\sum_k
\widetilde s_k
\end{equation}
The averaged quantities $\widetilde n_k$, $\widetilde s_k$ and
$\widetilde \phi_k$ are given by (see also \cite{braque})
\begin{eqnarray}\label{averaged}
\widetilde n_k =\widetilde n(\epsk) ={1\over\gav}\int_{-\infty}^{\infty}
n(e)
f_{av}\left({e-\epsk\over\gav}\right)de
\nonumber\\
\widetilde s_k =\widetilde s(\epsk) = {1\over\gav}\int_{-\infty}^{\infty}
s(e)
f_{av}\left({e-\epsk\over\gav}\right)de \nonumber\\
\widetilde \phi_k =\widetilde \phi (\epsk) = {1\over\gav}
\int_{-\infty}^{\infty}\phi(e) f_{av}\left({e-\epsk\over\gav}\right)de
\end{eqnarray}
With these quantities at our disposal we may express the first
order shell correction $\delta {\cal F}$ as \bel{delfree} \delta {\cal
F}(\Delta,T)=2\sum_k (\epsk\delta n_k -\Delta\delta\phi_k -T\delta s_k)
\end{equation}
with $\delta n_k=n_k-\widetilde n_k$, etc.
\begin{figure}[ht]
\centerline{{
\epsfxsize=16cm
\epsffile{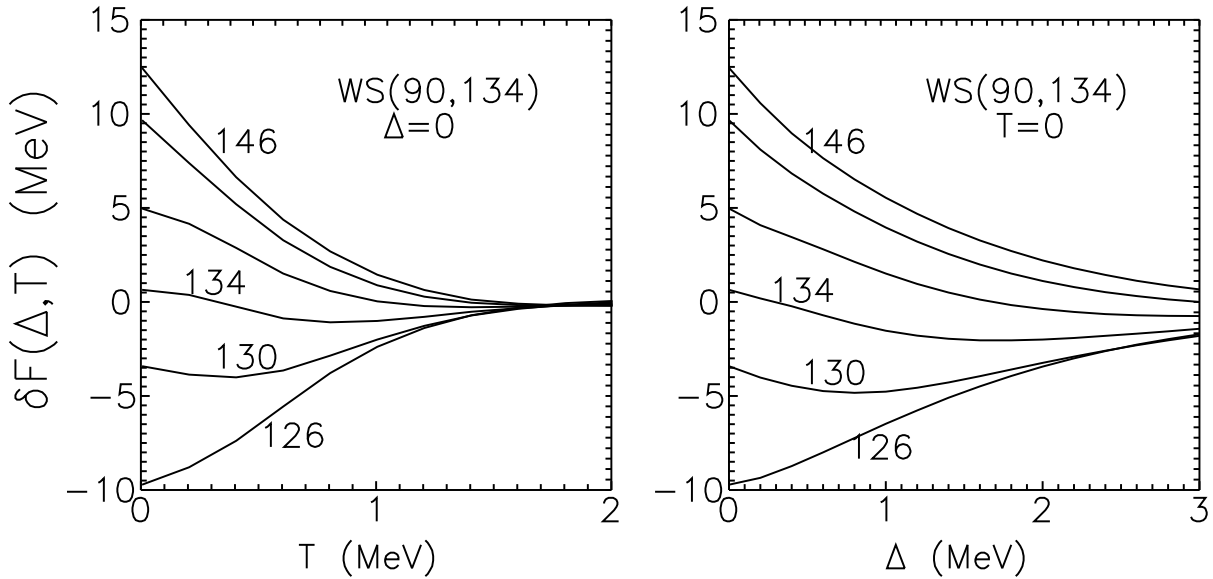}
}}
\caption{The shell correction to the free energy \protect\req{delfree}
for neutrons in the spherical Woods-Saxon potential
as function of temperature for $\Delta =0$
(left-hand-side) and as function of $\Delta$ for $T =0$ (right-hand-side).
The neutron numbers are indicated in the figure. The parameters of the
Strutinsky smoothing procedure are: averaging interval $\gav =8 \mev$,
order of the curvature polynomial: $M=6$.}
\label{fig1}
\end{figure}
In the vicinity of the Fermi energy the quantities $n_k$, $\phi_k$ and
$s_k$ are strongly varying functions of $\epsk$, but they behave rather
smoothly far away from it. As a consequence, the $\delta n_k$,
$\delta\phi_k$  and $\delta s_k$ differ substantially from zero only for
states close to the Fermi energy. This implies that the sums in
\req{delfree} are rather insensitive to a variation of their
limits. In particular, the problem of logarithmic convergence
in the solving gap equation does not appear in \req{delfree}.  Thus one need
not restrict the summation in \req{delfree} to the so called pairing window.

Fig.\ref{fig1} shows the dependence of the shell correction
\req{delfree} as function of temperature for $\Delta=0$ and as
function of $\Delta$ for $T=0$. It is seen that $\delta{\cal
F}(\Delta,T)$ decreases both with temperature as well as with
$\Delta$. The feature may be understood as follows. With growing $T$
or $\Delta$, the occupation numbers $n_k$ of the states close to the
Fermi energy  become more and more diffuse and thus resemble more
the average value $\widetilde n_k$. The same holds true for the
$s_k$ and $\phi_k$. Consequently, the absolute values of the $\delta
n_k$ etc. decrease and, hence, the sum \req{delfree} becomes
smaller. The temperature at which the shell structure disappears
is approximately equal to 1.5 $\mev$. Here, the shell correction
makes up only a few percent of its value at $T=0$. The decrease of
the shell correction with the pairing gap (at fixed temperature)
is less rapid. Even for $\Delta =2 \mev$ the shell correction still
amounts to a few dozen percent of its value at $\Delta=0$.

\subsubsection{Intrinsic response function for independent quasi-particles}
\label{subsubsec: chiintr}

The time-dependent $\mu\nu$-response functions for "intrinsic" or
nucleonic motion are to be calculated starting from the definition
\req{defrest}. To begin with we first address the model of independent
quasi-particles, discarding for the moment the modifications necessary
for "collisional damping", which will be discussed below. Moreover, we
want to concentrate on the $FF$ function. The results for the $FP$ and
the $PP$ response are given in the Appendix.

In the quasi-particle representation the (single-particle) operator
$\hat F$ can be written as
\bel{operator}
\hat F =\sum_k F_{kk}2v_k^2+\sum_{kj} F_{kj}\xi_{kj}(\alkp\alj+\alkbp\aljb)
+\sum_{kj} F_{kj} \eta_{kj}(\alkp\aljbp+\aljb\alk)
\end{equation}
where
\bel{etaxi}
\eta_{kj}\equiv u_kv_j+v_ku_j\,,\,\,\,\xi_{kj}\equiv u_ku_j-v_kv_j
\end{equation}
Inserting \req{operator} into \req{defrest} after somewhat lengthy
but straightforward calculations one gets
\begin{eqnarray}\label{chit}
\tilde\chi_{FF}(t)=
{-2\theta(t)\over\hbar}{\sum_{kj}}^{\pr}(n_k^T-n_j^T))\xi_{kj}^2\vert\fkj\vert^
2
\sin((E_k-E_j)t/\hbar)\nonumber\\ -{2\theta(t)\over\hbar}\sum_{kj}
(n_k^T+n_j^T-1)\eta_{kj}^2\vert\fkj\vert^2 \sin((E_k+E_j)t/\hbar)
\end{eqnarray}
Notice please, that in the first sum no diagonal components of the
$"\xi"$-terms survive. In the end this is due to the fact that the
corresponding terms of the operator $\foper$ commute with the
Hamiltonian \req{hprqp} and do not contribute to the response
function \req{chit}. That is why the first sum in \req{chit} is
marked by a prime. In the second sum of \req{chit} both diagonal
and non-diagonal components contribute. The Fourier transform of
\req{chit} lead to
\begin{eqnarray}\label{chiomlong}
\chi_{FF}(\om)={\sum_{kj}}^{\pr}(n_k^T-n_j^T)\xi_{kj}^2
\left({1\over{\hbar\om -(E_k-E_j)+i\varepsilon}}-{1\over{\hbar\om
+(E_k-E_j)+i\varepsilon}}
\right)\vert\fkj\vert^2
\nonumber\\
+\sum_{kj} (n_k^T+n_j^T-1)\eta_{kj}^2
\left({1\over{\hbar\om -(E_k+E_j)+i\varepsilon}}-{1\over{\hbar\om
+(E_k+E_j)+i\varepsilon}}
\right)\vert\fkj\vert^2
\end{eqnarray}
As it is seen from the last equation, the first line of
\req{chiomlong} does not contribute to the response function for
$T=0$ since in this case {\it all} $n_k^T=0$. For nonzero
temperature both sums of \req{chiomlong} do contribute. Please
observe that the second line only represents contributions from
excitations above $2\Delta$, as $E_k+E_j \geq 2\Delta$.
Conversely, as  $E_k-E_j$ may be arbitrarily small, contributions
to the truly low frequency region may only come from the first
line in \req{chiomlong}. Apparently they may built up only at
finite temperatures.

Eq.\req{chit} can be brought to the more compact form \bel{chitks}
\tilde\chi_{FF}(t)= {-\theta(t)\over \hbar}\{\sum_{k=j}\sum_{s\neq
s^{\pr}} +\sum_{k\neq j}\sum_{ss^{\pr}}\}
(n_{ks}^T-n_{\ksp}^T)\xi^2_{ks\ksp}\vert\fkj\vert^2
\sin\left({E_{ks}-E_{\ksp}\over\hbar}\right)t
\end{equation}
by introducing both positive as well as negative quasi-particle
energies \bel{eqpneg} E_{ks}=s{\sqrt{(\epsk
-\mu)^2+\Delta^2}}\,,\,{\rm with}\,\,s=\pm 1
\end{equation}
together with the corresponding amplitudes $u_{ks}$ and $v_{ks}$
\bel{uksvks}
u_{ks}={1\over \sqrt 2}\left(1-{\epsk- \mu\over E_{ks}}\right)^{1/2},\,\,\,\,
\,
v_{ks}={s\over \sqrt 2}\left(1+{\epsk- \mu\over E_{ks}}\right)^{1/2}
\end{equation}
The quantity $\xi_{ks\ksp}$ is defined similarly to \req{etaxi}
\bel{xi}
\xi_{ks\ksp}=u_{ks}\uksp-v_{ks}\vksp
\end{equation}
With the help of the quantities \req{eqpneg} and \req{xi} the Fourier
transform of \req{chitks} can be put into a form similar to the one at
$\Delta=0$,
\bel{chiom}
\chi_{FF}(\om)={1\over 2\hbar}\{\sum_{k=j}\sum_{s\neq s^{\pr}}
+\sum_{k\neq j}\sum_{ss^{\pr}}\}\xi^2_{ks\ksp}
{n_{ks}^T-n_{\ksp}^T\over {\hbar\om-(E_{ks}-E_{\ksp})+i\varepsilon}}
\vert\fkj\vert^2
\end{equation}
In the same way one may obtain the expressions for the other
components of the intrinsic response tensor, see the Appendix.
\subsubsection{Intrinsic response functions for collisional damping}
\label{subsubsec: damping}
To account for collisional widths of the quasi-particle states we
proceed similarly to the no paring case \cite{hofrep, ivhopaya}.
Let us explain details again at the example of the $FF$-response
function. The essence of the procedure is most easily seen after
first re-writing \req{chiom} in the form
\begin{eqnarray}\label{chiomgreen}
\chi_{FF}(\om)=
-{1\over 2\pi\hbar}\{\sum_{k=j}\sum_{s\neq s^{\pr}}
+\sum_{k\neq j}\sum_{ss^{\pr}}\}\xi^2_{ks\ksp}
\vert\fkj\vert^2\times\nonumber\\
\int_{-\infty}^{\infty}d\Om\,n^T(\Om)
\left[\varrho_{ks}(\Om){\cal G}^{(0)}_{\ksp}(\Om-\om-i\varepsilon)+
\varrho_{\ksp}(\Om){\cal G}^{(0)}_{ks}(\Om+\om+i\varepsilon)\right]
\end{eqnarray}
where ${\cal G}^{(0)}_{ks}(\om \pm i\varepsilon)$ is the Green function for
independent quasi-particles
\bel{greenks}
{\cal G}^{(0)}_{ks}(\om \pm i\varepsilon)={1\over{\hbar \om
    -E_{ks}\pm i\varepsilon}}
\end{equation}
The spectral density $\varrho_{ks}$ is related to ${\cal
G}^{(0)}_{ks}(\om \pm i\varepsilon)$ by
\bel{rhogreen}
\varrho_{ks}(\om)= i({\cal G}^{(0)}_{ks}(\om +i\varepsilon)
-{\cal G}^{(0)}_{ks}(\om -i\varepsilon))
\end{equation}

As argued in \cite{jehosi} the Green function ${\cal
G}^{{(0)}}_{ks}(\om \pm i\varepsilon)$ must be modified by
introducing a spreading of the quasi-particle width. This may be
achieved by introducing into the Green function a complex
self-energy $\self(\om\pm
i\varepsilon,\Delta,T)=\selfre(\om,\Delta,T)\mp i\Gamma(\om,\Delta,T)/2$
\bel{greenself} {\cal G}^{(0)}_{ks}(\om\pm
i\varepsilon)\rightarrow {\cal G}_{ks}(\om\pm i\varepsilon)=
{1\over{\hbar\om -E_{ks}-\selfre(\om)\pm i\Gamma(\om)/2}}
\end{equation}
The way of how $\Sigma$ is to be calculated in the presence of
pairing will be discussed in the next subsection. With ${\cal
G}_{ks}$ and $\varrho_{ks}$ given the response function
\req{chiomgreen} takes the form \bel{chiomkj} \chi_{FF}(\om)=
\{\sum_{k=j}\sum_{s\neq s^{\pr}}+\sum_{k\neq j}\sum_{ss^{\pr}}\}
\xi^2_{ks\ksp} \chi_{ks\ksp}(\om)\vert\fkj\vert^2
\end{equation}
where the amplitudes $\chi_{ks\ksp}(\om)$ are defined in the same
way as for the system without pairing, see \cite{ivhopaya}.
\bel{chikj}
\chi_{ks\ksp}(\om)=
\int_{-\infty}^{\infty}{d\Om\over 2\pi\hbar}\,n^T(\Om)
\left[\varrho_{ks}(\Om){\cal G}_{\ksp}(\Om-\om-i\varepsilon)+
\varrho_{\ksp}(\Om){\cal G}_{ks}(\Om+\om+i\varepsilon)\right]
\end{equation}
The only difference is that the energies $\epsk-\mu$ or $\epsj-\mu$ in
eq.(4.2) of \cite{ivhopaya} are replaced by $E_{ks}$ or $E_{js^{\pr}}$.

The folding integral in \req{chikj} is computed by means of the
residue theorem closing the integration contour in the lower half
of the frequency plane. The residues which contribute to the
integral \req{chikj} are that of ${\cal G}_{ks}(\Om\pm
i\varepsilon)$ (mind \req{rhogreen}) and $n^T(\Om)$. In case of
the frequency independent width $\Gamma=\Gamma(\Delta,\Delta,T)$ used
here (see also sect.\ref{subsec: gamma} below) the poles of ${\cal
G}_{ks}(\Om\pm i\varepsilon)$ are given by \bel{gpoles}
\hbar\Om_{\pm}=E_{ks}\mp i \Gamma/2
\end{equation}
and the poles of $n^T(\Om)$  are given by the Matsubara
frequencies \bel{npoles} \hbar\Om_{n}=\pm i \pi T (2n+1), \qquad
n=0,1,2,...
\end{equation}
The computation of the folding integral \req{chikj} with a
frequency dependent width $\Gamma(\om,T)$ without pairing is
discussed in detail in \cite{ivhopaya}.
\subsection{The quasi-particle strength function}
\label{subsec: gamma}
The Green function ${\cal G}_{ks}$ \req{greenself} contains the
frequency dependent self-energy $\self$. Let us first focus our
attention on its imaginary part $-Im \self = \Gamma/2$, which we
will estimate from the collision term of the Boltzmann equation
for the scattering of two quasi-particles. Its standard form can
be found in the review of Baym and Pethik \cite{baympe} for
ordinary particles and in \cite{bhatta1, bhatta2} for Bogolyubov
quasi-particles. As is well known, for $\Delta=0$ and for small
excitations the width (decay rate) $\Gamma_d(\hbar\om)$ can be
written as \bel{gammaipm}
\Gamma_d(\hbar\om)={1\over\Gamma_0}[(\hbar\om -\mu)^2+\pi^2T^2]\,.
\end{equation}
For Bogolyubov quasi-particles no analytic expression for this
decay rate $\Gamma_d(\hbar\om)$ is available, rather this quantity has
to be computed numerically. In the present work we followed the
publications \cite{jehosi, sijecop} where $\Gamma_d(\hbar\om)$ was
expressed as \bel{gammaint} \Gamma_d(\hbar\om)={2\over
\Gamma_0}\int \int \int d\eps_2 d\eps_3 d\eps_4
\delta(\hbar\om+{\cal E}_2+{\cal E}_3+{\cal
E}_4)[n^{T}_2n^{T}_3n^{T}_4+ (1-n^{T}_2)(1-n^{T}_3)(1-n^{T}_4]
\end{equation}
Here, the $n^{T}_i$ represent the thermal equilibrium
distribution $n^T({\cal E}_i)= (1+\exp({\cal E}_i/T))^{-1}$ for
the quasi-particles, the energies of which are defined as
\bel{cale}
{\cal E}_k=E_k{\epsk -\mu\over\vert\epsk
-\mu\vert}
\end{equation}
They are positive (negative) for states arising from single
particle states above (below) the Fermi energy. The $\Gamma_0$
depends on the averaged matrix element of the residual interaction, the
effective mass $m^*$ at the
Fermi energy and a cut-off momentum $q_c$:
\bel{gammazero}
{1\over\Gamma_0}=\left({m^*\over 2\pi\hbar}\right)^3{q_c\over p_F}
{\left\arrowvert\overline{\langle 34\vert g \vert
12\rangle}\right\arrowvert^2}
\end{equation}
In deriving \req{gammaint} an expansion in the momenta  around the
Fermi momentum was used, and the $q_c$ is an upper limit for the
momentum transfer involved in the scattering, which is assumed to
be small compared to the Fermi momentum $q_c (\leq p_F)$, see
\cite{baympe}. Therefore, the result \req{gammaint} is valid for
small excitations only.  Fortunately, it is this region around the
Fermi energy which gives the main contribution to the response
functions. In case $\Gamma_d$ is needed also for larger energies a
correction becomes necessary. This problem is not specific to
pairing, it already exists for the zero gap limit where it was
solved by introducing one more parameter, see \cite{jehosi,
sijecop}. Generalizing this procedure to the case of pair
correlations, we may assume the final form of decay width to be
given by \bel{gamma} \Gamma={\Gamma_d\over{1+\Gamma_d
\Gamma_0/c^2}}
\end{equation}
with the cut-off parameter $c$ (in energy) chosen to be
independent of $\Delta$ and equal to the value $c=20 \mev$ given
already in \cite{jehosi}.   Likewise, for the $\Gamma_0$ of
\req{gammazero} we take the same constant as before, namely
$\Gamma_0=33 \mev$ (see \cite{jehosi}).

\begin{figure}[t]
\centerline{{\epsfxsize=10cm \epsffile{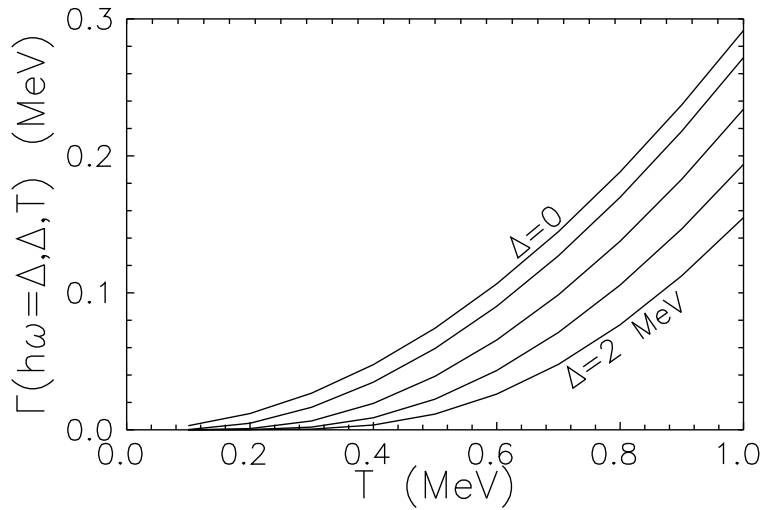}}}
\caption{The collisional width \protect\req{gamma} taken at
the Fermi energy ($\hbar\om =\Delta$) as function of the temperature.
Different curves correspond to different values of pairing gap
$\Delta$ indicated in the figure.}
\label{fig2}
\end{figure}

The computation of the folding integral \req{chiomgreen} with the
frequency dependent width $\Gamma$ requires the knowledge of
$\Gamma$ in the whole range of $\om$ from zero to infinity. This
turns out to be a rather difficult task. Already a calculation of
the $\Gamma_d$ through \req{gammaint}) would involve  a double
integral, which in the end would be too time consuming. To reduce
the computation time to manageable level we chose to work with a
constant $\Gamma$, taken at the Fermi energy. This is to say that
in the present numerical evaluations we replaced
$\Gamma(\hbar\om)$ by $\Gamma=\Gamma(\hbar\om=\Delta)$. This
quantity is shown in Fig.\ref{fig2} as function of temperature
for a few values of the pairing gap. As might be expected on
general grounds, for $T=0$ the collisional width \req{gamma} is
zero for all $\hbar\om \le 2\Delta$ (incidentally, in
\cite{jehosi} it was found to vanish in the regime $\hbar\om \leq
3\Delta$).  It is also seen that $\Gamma$ is the smaller the
larger is $\Delta$. This effect is especially strong for small
temperatures, say  $T\leq 0.5 \mev$.

Once the $\Gamma(\hbar\om,\Delta,T)$ is known the real part of
$\self$ is given by Kramers-Kronig relation. For systems without
pairing it was shown analytically \cite{ivhopaya} that $\selfre
\sim \hbar\om - \mu$. The same is true in the presence of pairing.
The numerical results exhibit the $\selfre$ to vanish at the Fermi
energy and to be very small in its vicinity. Within the
approximations used here, namely to take the width at the Fermi
energy where $\selfre=0$, the real part of the
self-energy does not contribute to the response functions.

Finally, we should like to mention some shortcomings of the
derivation of \req{gammaint} given in \cite{jehosi}, which so far
we have not been able to improve, but which for the present works.
1) In the vicinity of the critical temperature $T_c$, where the
pairing gap is small, not only the scattering but also the
coalescence and decay of quasi-particles should be taken into
account, see \cite{bhatta1}. 2) For small temperatures, where the
pairing gap is large, the introduction of the energies ${\cal
E}_k$, instead of quasi-particle energies $E_k$ --- which make the
collision integral resemble that for ordinary particles as closely
possible --- does not appear to be well justified; moreover, one
would think that the decay rate should also be influenced by the
amplitudes $u_k^2$ and $v_k^2$. It is very likely, of course, that
such finer details will not modify our final results too much, as
we are using drastic approximations to the width anyway. More
detailed investigations are required to clarify these points.
\section{Average particle number conservation}
\label{sec: spurious}

In this chapter we address the problem of particle number
conservation. As one knows, violation of any symmetry may lead to
spurious states. Their existence in the present model may be seen
with the help of the $FF$-response function, for instance. Its
poles are associated with the energies of the systems' excited
states. As it is seen from \req{chiomlong} some of the poles
correspond to pair excited states $\alkp\alkbp\vert 0\ket$ with
energies $2E_k$. The number of these states is equal to the number
of terms in the sums over $k$ or $j$ in \req{chiomlong}), one more
than the right value, as argued in \cite{hogaasen}. This extra
states is the spurious one. Fortunately, it is possible to
construct response functions where such contributions are removed,
as shall be explained now.

\subsection{Number conserved response}
\label{subsec: numconres}

\subsubsection{Modified intrinsic response}
\label{subsubsec: mointresp}

For the independent quasi-particle model applied to the
quasi-static picture, it is the chemical potential $\mu$ which
guarantees the conservation of the average particle number
determined by eq.\req{partnum}. However, the dynamical density
matrix will differ from the quasi-static density $\rho_{\rm qs}$.
Therefore, in  a time dependent picture the average particle
number $\bra N\ket_t$ does not automatically coincide with $N$.
Consequently, spurious contributions proportional to $\delta
\bra\hat N\ket_t$ might appear. Following \cite{abrstr}and
\cite{abrosim}, one may correct this drawback {\it requiring as a
constraint} $\delta\bra\hat N\ket_t=0$ to hold true, thus warranting
that the average particle number is conserved also in dynamical
sense.

To implement such a constraint into the theory let us recall first
that the Hamiltonian \req{hprqp} depends on the shape
parameter $Q$, the chemical potential $\mu$ and
$\Delta$. The change with time of the average of some operator, say
$\bra\foper\ket_t$, will generally depend on the variation of all three
parameters $Q$, $\mu$ and $\Delta$. To evaluate the deviation of
$\bra\foper\ket_t$ one may use the set of equations \req{delfcoll}
which in the present case read:
\bel{delfom}
\delta \bra \hat F \ket_{\om}=-\chi_{FF}(\om)\delta Q(\om)+
\chi_{FN}(\om)\delta\mu(\om)+\chi_{FP}(\om)\delta\Delta(\om)
\end{equation}
\bel{delpom}
\delta \bra (\hat P+\hat P^\dagger)\ket_{\om}=-\chi_{PF}(\om)\delta Q(\om)
+\chi_{PN}(\om)\delta\mu(\om)+\chi_{PP}(\om)\delta\Delta(\om)
\end{equation}
A similar relation must hold true for $\delta \bra\hat N\ket_{\om}$
\bel{delnom} \delta \bra\hat N\ket_{\om}=-\chi_{NF}(\om)\delta Q(\om)+
\chi_{NN}(\om)\delta\mu(\om)+\chi_{NP}(\om)\delta\Delta(\om)
\end{equation}
As mentioned previously, the expressions for the $FN$-,
$NP$-response functions may be found in the Appendix. The
condition $\delta \bra\hat N\ket_{t}=0$ or $\delta \bra\hat
N\ket_{\om}=0$ together with \req{delnom} allows one to express
$\delta\mu(\om)$ in terms of $\delta Q(\om)$ and $\delta\Delta(\om)$
\bel{delmu} \delta\mu(\om)={\chi_{NF}(\om)\over\chi_{NN}(\om)}\delta
Q(\om) -{\chi_{NP}(\om)\over\chi_{NN}(\om)}\delta\Delta(\om)
\end{equation}
By substituting \req{delmu} into \req{delfom} and \req{delpom} the
latter two equations turn into
\begin{eqnarray}\label{delfdelp}
\delta \bra \hat F\ket_{\om}=
-\widehat\chi_{FF}(\om)\delta Q(\om)+\widehat\chi_{FP}(\om)\delta\Delta(\om)
\nonumber\\
\delta\bra (\hat P+\hat P^\dagger)\ket_{\om}=
-\widehat\chi_{PF}(\om)\delta Q(\om)+\widehat\chi_{PP}(\om)\delta\Delta(\om)
\end{eqnarray}
In this way we get new response functions $\widehat\chi_{FF}(\om)$,
$\widehat\chi_{FP}(\om)$ and
$\widehat\chi_{PP}(\om)$ which account for the conservation
of the average particle number during a dynamical process. They may be
constructed from the former one by the simple relations:
\bel{chiff}
\widehat\chi_{FF}(\om)=\chi_{FF}(\om)
-{\chi_{FN}(\om)\chi_{NF}(\om)\over\chi_{NN}(\om)}
\end{equation}
\bel{chifp}
\widehat\chi_{FP}(\om)=\chi_{FP}(\om)
-{\chi_{FN}(\om)\chi_{NP}(\om)\over\chi_{NN}(\om)}
\end{equation}
\bel{chipp}
\widehat\chi_{PP}(\om)=\chi_{PP}(\om)
-{\chi_{PN}(\om)\chi_{NP}(\om)\over\chi_{NN}(\om)}
\end{equation}

\subsubsection{Modified collective response}
\label{subsubsec: mocollresp}
We may now use the information contained in \req{chiff}-\req{chipp},
\req{appaqq}, and \req{paircoco}
to construct the collective response tensor \req{collresp}.
The expressions for the component of
$\mbox{{\boldmath $\chi$}}_{\rm coll}(\om)$ can be
obtained directly from the form given in \req{collresp}. One gets
\bel{chiqqcoll}
\chi_{FF}^{\rm coll}(\om)=
{{\kappa_{Q}[(\kappa_{\Delta}+\hat\chi_{PP})\hat\chi_{FF}-
\hat\chi_{FP}\hat\chi_{FP}]}
\over
{(\kappa_{Q}+\hat\chi_{FF})(\kappa_{\Delta}+\hat\chi_{PP})-
\hat\chi_{FP}\hat\chi_{PF}}}
\end{equation}
\bel{chiqpcoll}
\chi_{FP}^{\rm coll}(\om)=
{{\kappa_{Q}\kappa_{\Delta}\hat\chi_{FP}}
\over
{(\kappa_{Q}+\hat\chi_{FF})(\kappa_{\Delta}+\hat\chi_{PP})-
\hat\chi_{FP}\hat\chi_{PF}}}
\end{equation}
and
\bel{chippcoll}
\chi_{PP}^{\rm coll}(\om)=
{{\kappa_{\Delta}[(\kappa_{Q}+\hat\chi_{FF})\hat\chi_{PP}-
\hat\chi_{FP}\hat\chi_{FP}]}
\over
{(\kappa_{Q}+\hat\chi_{FF})(\kappa_{\Delta}+\hat\chi_{PP})-
\hat\chi_{FP}\hat\chi_{PF}}}
\end{equation}
The collective $Q$-vibrational frequency $(\Om_1)$ is defined by
the lowest pole of collective response function \req{chiqqcoll} or
by the equation \bel{seqequ}
{(\kappa_{Q}+\hat\chi_{FF}(\Om_1))(\kappa_{\Delta}+\hat\chi_{PP}(\Om_1))-
\hat\chi_{FP}(\Om_1)\hat\chi_{PF}(\Om_1)}=0.
\end{equation} It is not difficult to convince oneself that \req{seqequ}
may as well be interpreted as the secular equation of the system
\req{delfom}-\req{delnom}, provided one obeys the subsidiary
condition $\delta \bra\hat N\ket_{\om}=0$ as well as the $Q$- and
$\Delta$-components of self-consistency conditions \req{manyself}.
In case we are interested in vibrations around a minimum of the
potential energy, both with respect to $Q$ and $\Delta$, we may
identify $Q_m$ with $Q_0$, $\Delta_m$ with $\Delta_0$. Thus the $Q$-
and $\Delta$-components of $q(\om)$ will coincide with $\delta Q(\om)$
or $\delta \Delta(\om)$ and \req{manyself} will attain the form
\bel{subsid} \delta\bra\hat F\ket_{\om}=\kappa_Q\delta Q(\om)\,,\quad
\delta \bra\hat P+\hat P^\dagger\ket_{\om}={2\over G}\delta \Delta(\om)
\end{equation}
Excluding $\delta\bra\hat F\ket_{\om}$,
$\delta \bra\hat P+\hat P^\dagger\ket_{\om}$ and $\delta\bra\hat
N\ket_{\om}$ from the system
\req{delfom}-\req{delnom} with the help of \req{subsid} together with
$\delta\bra\hat N\ket_{\om}=0$ one will get the following system of equations
\begin{eqnarray}\label{system}
-(\chi_{FF}(\om)+\kappa_Q)\delta Q(\om)&+\chi_{FN}(\om)\delta\mu(\om)
&+\chi_{FP}(\om)\delta\Delta(\om)=0\nonumber\\
-\chi_{NF}(\om)\delta Q(\om)&+\chi_{NN}(\om)\delta\mu(\om)
&+\chi_{NP}(\om)\delta\Delta(\om)=0\\
-\chi_{PF}(\om)\delta Q(\om)&+\chi_{PN}(\om)\delta\mu(\om)
&+(\chi_{PP}(\om)-2/G)\delta\Delta(\om)=0\nonumber
\end{eqnarray}
The secular equations is then given by putting the determinant of
the matrix behind this set equal to zero. Indeed, this leads to
the same result as in \req{seqequ}. Below, this fact shall be
exploited further for special cases.
\subsection{RPA at zero thermal excitations}
\label{subsec: rpazero}
It should be worth while to compare our results with those
obtained in previous treatments at zero temperature. Using the
expressions for the intrinsic response functions given in the
Appendix for $T=0$ one derives from \req{system} the following
secular equation \bel{determin}
\begin{array}{ccccccc}
&\Bigg\arrowvert&\sum_{kj} {2\eta^2_{kj}E_{kj}\vert F_{kj}\vert^2\over
E_{kj}^2-\hbar^2\om^2}+{\kappa_Q}&
4\Delta\sum_k {\eta_{kk}F_{kk}\over4E_k^2-\hbar^2\om^2}&
4\Delta\sum_k {\xi_{kk}F_{kk}\over4E_k^2-\hbar^2\om^2}&\Bigg\arrowvert&\\
{\rm Det}_{LRT}(\om)\equiv
&\Bigg\arrowvert&4\Delta\sum_k {\eta_{kk}F_{kk}\over4E_k^2-\hbar^2\om^2}&
4\Delta\sum_k {\eta_{kk}\over 4E_k^2-\hbar^2\om^2}&
4\Delta\sum_k {\xi_{kk}\over 4E_k^2-\hbar^2\om^2}&\Bigg\arrowvert&=0\\
&\Bigg\arrowvert&4\Delta\sum_k {\xi_{kk}F_{kk}\over4E_k^2-\hbar^2\om^2}&
4\Delta\sum_k {\xi_{kk}\over4E_k^2-\hbar^2\om^2}&
\sum_k {4E_k\xi_{kk}^2\over4E_k^2-\hbar^2\om^2}-{2\over G}&\Bigg\arrowvert&
\end{array}
\end{equation}
This equation may be compared with the secular equations obtained
in \cite{besszy, solo} and used for investigations of vibrational
states in medium and heavy nuclei. The determinants of
\cite{besszy} and \cite{solo} differ from each other by a constant
multiplier so that the collective frequencies are the same. The
one obtained in \cite{solo} (let us denote it as ${\rm
Det}_{RPA}(\om)$) writes \bel{detsolo}
\begin{array}{ccccccc}
&\Bigg\arrowvert&\sum_{kj} {\vert F_{kj}\vert^22\eta_{kj}^2(E_k+E_j)\over
(E_k+E_j)^2-\om^2}+{1\over k}&
\om\sum_k {\eta_{kk}F_{kk}\over 4E_k^2-\om^2}&
2\sum_k {\xi_{kk}\eta_{kk}F_{kk}\over 4E_k^2-\om^2}&\Bigg\arrowvert&\\
{\rm Det}_{RPA}(\om)\equiv&\Bigg\arrowvert
&2\om\sum_k {\eta_{kk}F_{kk}\over 4E_k^2-\om^2}&
\sum_k {2E_k\over 4E_k^2-\om^2}-{1\over G}&
\om\sum_k {\xi_{kk}\over 4E_k^2-\om^2}&\Bigg\arrowvert&\\
&\Bigg\arrowvert&4\sum_k {\xi_{kk}\eta_{kk}F_{kk}\over4E_k^2-\om^2}&
\om\sum_k {\xi_{kk}\over 4E_k^2-\om^2}&
\sum_k {2E_k\xi_{kk}^2\over 4E_k^2-\om^2}-{1\over G}&\Bigg\arrowvert&
\end{array}
\end{equation}
To establish a relation between \req{determin} and \req{detsolo}
let us first recall that in \cite{besszy, solo} vibrations around
the minimum of the potential energy were considered. In this case,
because of the linearization involved, the pairing gap may be
defined through the solution of the gap equation (corresponding to
the minimum) and one may replace $1/G$ by $\sum_k 1/2E_k$.
Consequently, one has \bel{substit} \sum_k {2E_k\over
4E_k^2-\om^2}-{1\over G}= {\om^2\over2\Delta}\sum_k
{\eta_{kk}\over 4E_k^2-\hbar^2\om^2}
\end{equation}
Considering this relation, the determinants \req{determin} and
\req{detsolo} look very similar. Indeed, by explicit calculation
one may verify the following relation \bel{relation} {\rm
Det}_{RPA}(\om)={\om^2\over(4\Delta)^2}{\rm Det}_{LRT}(\om)
\end{equation}
It is clear from \req{relation} that our secular equation
\req{determin} defines the nontrivial part of the secular equation
of \cite{besszy, solo}. The spurious mode corresponding to $\om=0$
is not present among the solutions of \req{determin}, which are
the same as those found in \cite{besszy, solo}.

Later on we are going to evaluate transport coefficients. They are
largely determined by the strength distributions associated to the
frequency dependence of response functions, like $\chi_{FF}(\om)$
and $\hat\chi_{FF}(\om)$. Our numerical computations show that the
shift in the strength distribution due to correction terms like
$\chi_{FN}(\om) \chi_{FN}(\om)/\chi_{NN}(\om)$  is small. This
implies that particle number conservation has only little
influence on the transport coefficients. The correction terms
mostly cause shifts in the positions of the peaks of (the
imaginary parts of) the response function, whereas the widths of
the latter remain almost unchanged. As a consequence, the friction
coefficients deduced from $\chi_{FF}(\om)$ and
$\hat\chi_{FF}(\om)$ are almost identical. Not much difference is
seen is also in the mass parameter and the local stiffness.
\subsubsection{Pairing without shape oscillations}
\label{subsubsec: pairwi}
For a comparison with the standard pairing theory let us consider
the particular case where the coupling between $Q$-mode and
$\Delta$-mode may be neglected ($\hat\chi_{FP}(\om)=0$). In this
case the secular equation \req{seqequ} separates into individual
ones for the $Q$- and the $\Delta$-mode. The latter can be written
as \bel{parsequ} \widehat\chi_{PP}(\om)-{2\over
G}=\chi_{PP}(\om)-{2\over G}
-{\chi_{PN}(\om)\chi_{NP}(\om)\over\chi_{NN}(\om)}=0
\end{equation}
Replacing $2/G$ in the same way as above by $\sum_k 1/E_k$ and
using the explicit expressions for the response functions,
eq.\req{parsequ} is transformed to \bel{squared}
(\hbar^2\om^2-4\Delta^2) \left(\sum_k {1\over
E_k(4E_k^2-\hbar^2\om^2)}\right)^2 = \left(\sum_k {2(\epsk
-\mu)\over E_k(4E_k^2-\hbar^2\om^2)}\right)^2
\end{equation}
After taking the square root eq.\req{squared} is reduced to
\bel{quasiboson} \sum_k {\sqrt{\hbar^2\om^2-4\Delta^2}\pm2(\epsk
-\mu) \over E_k(4E_k^2-\hbar^2\om^2)} = 0
\end{equation}
This equation is identical to that obtained in \cite{hogaasen,
cushar} within quasi-boson approximation. In this sense essential
results of the theory of vibrational states in paired systems are
reproduced by our linear response approach.

\bigskip\bigskip
\section{Numerical results}
\label{sec: results}
In this section we are going to present numerical results for
response functions and transport coefficients. We concentrate on
deformations along the fission path of the nucleus $^{224}Th$,
which has been under investigation before. Indeed, present one is
an extension of previous publications \cite{yaivho, ivhopaya}
where temperatures above $T=1 MeV$ had been considered with
pairing discarded. Like in \cite{ivhopaya} we use the independent
particle Hamiltonian based on the deformed Woods-Saxon potential
with the nuclear shape parameterized in terms of Cassini ovaloids
\cite{pash71}. The Cassini ovaloids are defined by rotating the
curve \bel{ovals} \rho(z, \epsilon)=R_0\left[\sqrt{a^4+4\epsilon
z^2/R_0^2}-z^2/R_0^2-
     \epsilon\right]^{1/2}
\end{equation}
around the $z$-axis, with $z$ and $\rho$ being cylindrical coordinates.
The constant $a$ is determined from volume conservation, implying that the
family of shapes (\ref{ovals}) depends only on one deformation parameter
$\epsilon$. As is easily recognized from (\ref{ovals}) the value of
$\epsilon =0$ corresponds to a sphere. For $0 < \epsilon < 0.4$
the form \req{ovals} resembles very much that of a spheroid with the ratio of
the axes given by
\begin{equation}
\label{ratio}
{\rm{shorter\,\,\,axes}\over \rm{longer\,\,\,axes}}=
    {{1-2\epsilon /3}\over{1+\epsilon /3}}
\end{equation} At $\epsilon \approx 0.5$ a neck appears and at
$\epsilon =1.0$ the nucleus separates into two fragments. Like in
\cite{ivhopaya}, instead of $\epsilon$  we will use the parameter
$Q=R_{12}/2R_0$, which measures the distance $R_{12}$ between the
left and right center of masses divided by the diameter $2R_0$ of
the sphere of identical volume. Besides having a simple physical
meaning, the $Q=R_{12}/2R_0$ allows one to relate to other shape
parameterizations of deformed potentials.

In principle, the pairing gap is different for neutrons and
protons. Thus, for quantitative studies one should introduce gaps
for protons and neutrons as two independent dynamical variables.
However, in the present work we would like to concentrate more on
questions of principle nature, as a general investigating of the
effects of pairing on transport coefficients.  Thus we feel
justified to compute the response functions putting $\Delta^p
=\Delta^n=\Delta$ and to use this $\Delta$ as a collective parameter
common for protons and neutrons. This will simplify also the
presentation of our results.

\begin{figure}[p]
\centerline{{\epsfxsize=16cm \epsffile{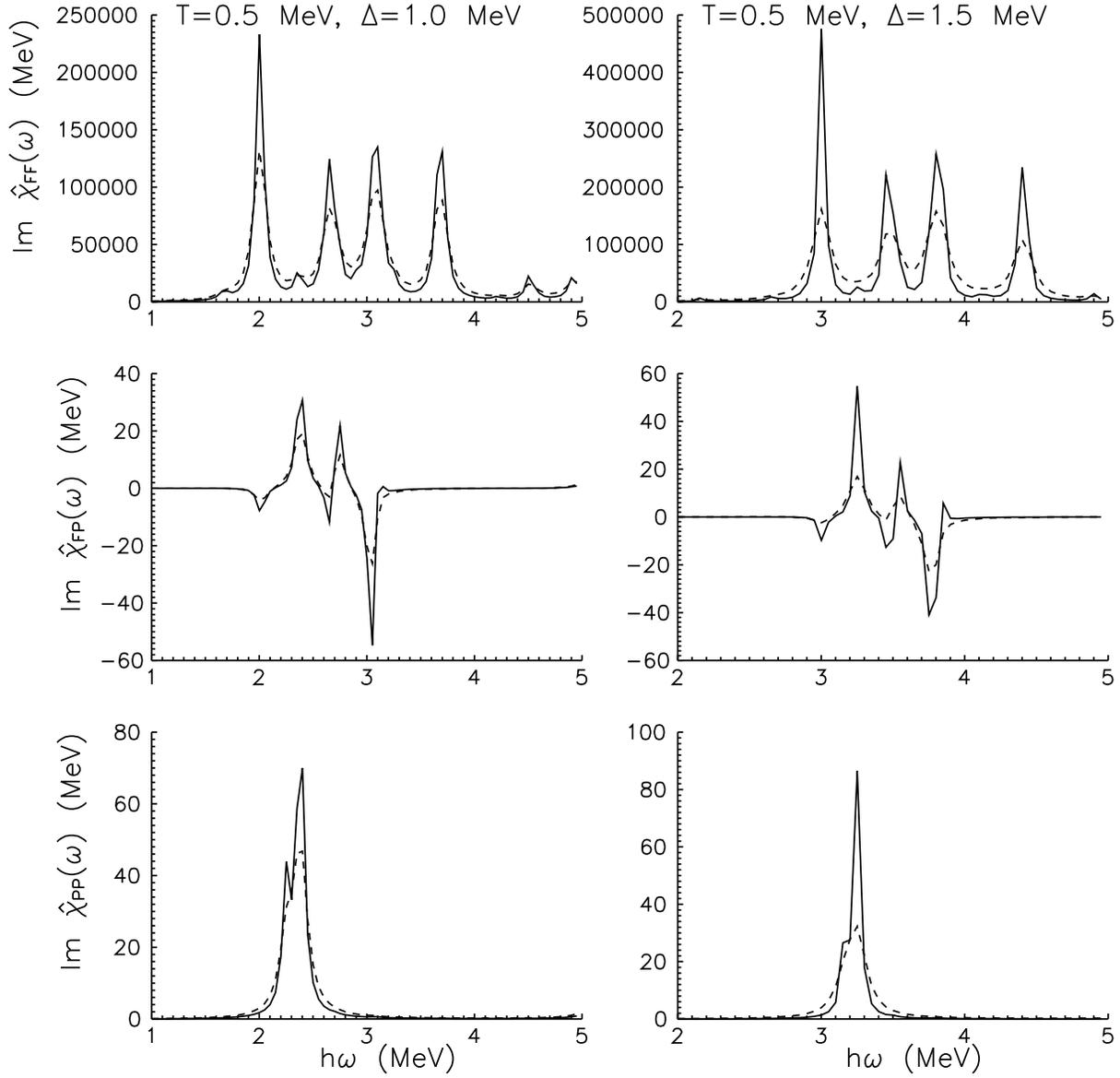}}}
\caption{The frequency dependence of the imaginary part of the
intrinsic response functions
\protect\req{chiff}-\protect\req{chipp}. The computations were
done with the spherical Woods-Saxon potential for the nucleus
$^{224}Th$. The values of the temperature $T$ and $\Delta$ are
indicated in the figure. The dashed and solid curves correspond to
different choices of the collisional widths, namely
\protect\req{gammaipm} and \protect\req{gammaint}, respectively
(together with \protect\req{gamma}). } \label{fig3}
\end{figure}

\subsection{Information from the intrinsic response}
\label{subsec: numchiin}

We begin looking at the intrinsic response functions
$\chi_{\mu\nu}$, as the most basic quantity from which all local
transport properties may be deduced, once the static energy is
known. For instance, it allows for an immediate calculation of
inertia and friction within the zero frequency limit, which for
not too small temperature may give reasonably accurate results, in
particular in the version suggested in \cite{ivhopaya}. Moreover,
this response function is a basic element in the construction of
the collective response function, which in turn allows one to
evaluate the implication of self-consistency on transport
coefficients. It is thus of great interest to look first at the
effects pairing has on the $\chi_{\mu\nu}$. In Fig.\ref{fig3}
we show the dissipative parts of the susceptibilities given in
\req{chiff}-\req{chipp} as function of frequency, for $T=0.5 MeV$
and two values of $\Delta$, for spherical configurations. The
functions presented in all three parts of the figure exhibit a
narrow peak structure. These peaks represent the excitations the
system would have if there were no correlations due to collective
motion.

Let us first look at the effects the collisional width
$\Gamma(\hbar\om=\Delta,\Delta,T)$ has on width and height of the
peaks in the response functions. To clarify this influence, and
consequently the one on the transport coefficients, in
Fig.\ref{fig3} two sets of curves are  plotted: One computed with
$\Gamma(\hbar\om=\Delta,\Delta =0,T)$ (dashed curves) and another
one with the $\Gamma(\hbar\om=\Delta,\Delta,T)$ given by
eqs.\req{gamma} and \req{gammaint}. As it is seen from the figure,
the peaks computed with a finite $\Delta$ are higher and more
narrow as compared with those for the case $\Delta=0$.  This is to
be expected since, as it is seen from Fig.\ref{fig2} for fixed
temperature, $\Gamma(\hbar\om=\Delta,\Delta,T)$ is the smaller the
larger is $\Delta$. For fixed $\Delta$ the width
$\Gamma(\Delta,\Delta,T)$ increases with temperature and, as the
computations demonstrate, the response functions become smoother.

Next we examine the influence of the spectrum of the individual
excitations. As compared to the unpaired case, the response
functions shown in Fig.\ref{fig3} exhibit a totally different
structure: The lowest peak of considerable strength occurs at or
above the frequency $\hbar\om\approx 2\Delta$, and the strength in
the region $\hbar\om\leq 2\Delta$ is very small. In case of the
$PP$-response, for instance, the position of the peaks is defined
by the zeros of eq.\req{quasiboson}. The two peaks seen at the
bottom part of Fiq.\ref{fig3} represent the contributions from
the lowest solutions of eq.\req{quasiboson} for neutrons and
protons. For a spherically symmetric potential the lowest peaks
are well isolated from the rest. This is because the positions of
the peaks of the PP-response function is mainly determined by the
energy  2$E_k$ of pair excitations. This distance is especially
large for spherical configurations, simply because of the large
degeneracies of the single-particle states. For deformed shapes
the distance to the next peaks becomes much smaller.

\begin{figure}[ht]
\centerline{{\epsfxsize=16cm \epsffile{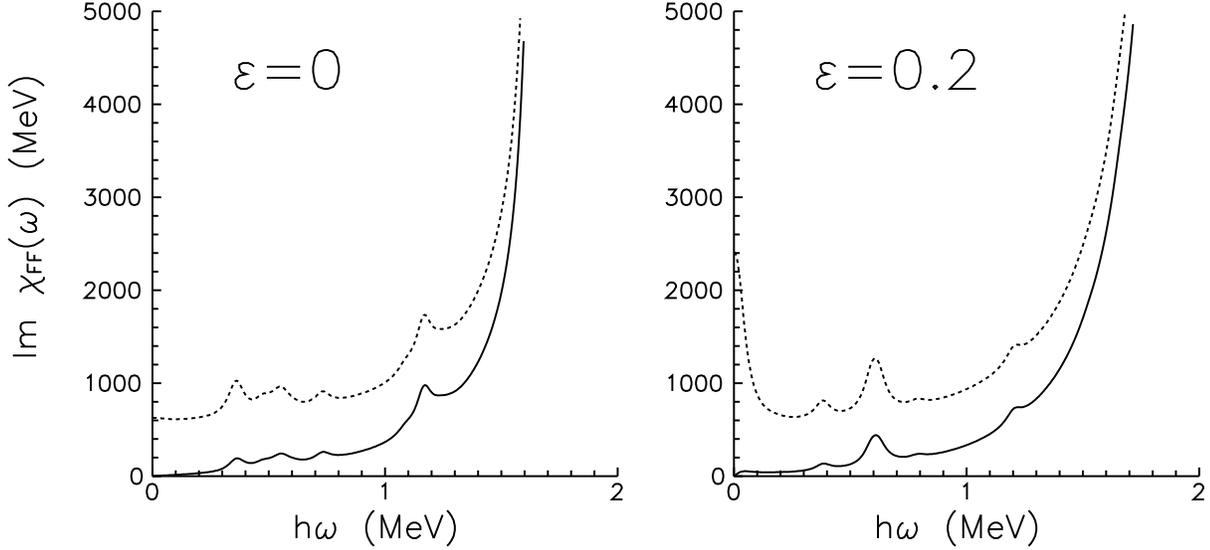}}}
\caption{The frequency dependence of the imaginary parts of the response
(solid curves) and correlation (dotted curves) functions. The computations are
done
with the spherical (left) and deformed (right, $\eps=0.2$) potentials. }
\label{fig4}
\end{figure}

Obviously, these features will have great impact on transport
coefficients. This is most apparent for the friction coefficients
defined in zero frequency limit: They will be the smaller the
larger is $\Delta$. This property also shows up  more or less
strongly for the self-consistent calculation. Let us therefore
analyze the structure of the response functions at small
excitations in more detail. It is easily recognized that it may be
traced back to the various contributions to formula
\req{chiomlong} (or the corresponding one resulting after
collisions are taken into account).  As we will see, the
contributions from the two lines behave very differently,
depending very sensitively on temperature. Let first look at the
case of finite temperatures. Here, the first line of
\req{chiomlong}, resulting from the terms with $\vert E_k-E_j \vert \leq
2\Delta$,  will have finite contributions even in the region
$\hbar\om\leq 2\Delta$. However, on the scale used in the figure,
they are very small. Fig.\ref{fig4} shows the (dissipative
part of the) response function in the interval below $2\Delta$ in
more detail, exhibiting clearly the existence of such
contributions.

Since around $\om =0$, the region which is of special interest,
this dissipative part has to vanish exactly, it is more
instructive to use the {\it correlation function}
$\psi^{\dpr}_{\mu\nu}(\om)$, related to the former by the
fluctuation dissipation theorem (for the nucleonic degrees of
freedom), see \cite{hofrep}. One has to replace the commutator of
\req{defrest} by an anti-commutator and must account for the
unperturbed averages relative to which the deviations $\delta
F_\mu = F_\mu - \bra F_\mu \ket$ of the operators are to be
defined, \bel{corfunt} \widetilde\psi^{\dpr}_{\mu\nu}(t)= {1\over
2}( \bra [\hat F_{\mu}^{I}(t),\hat F_{\nu}]_+\ket
    - \bra \hat F_{\mu}\ket \bra\hat F_{\nu}\ket )
\end{equation}
Calculating the averages in \req{corfunt} in the same way as for the
response functions, for independent quasi-particles the
Fourier transform of the $FF$-correlation function turns out to be
\begin{eqnarray}\label{psiomlong}
\psi_{FF}(\om)=-{\sum_{kj}}^{\pr}n_k^T(1-n_j^T)\xi_{kj}^2
\left({1\over{\hbar\om -E_{kj}^-+i\varepsilon}}+{1\over{\hbar\om
+E_{kj}^-+i\varepsilon}}
\right)\vert\delta\fkj\vert^2
\nonumber\\
-\sum_{kj} [(n_k^Tn_j^T+(1-n_k^T)(1-n_j^T)]\eta_{kj}^2
\left({1\over{\hbar\om -E_{kj}^++i\varepsilon}}+{1\over{\hbar\om
+E_{kj}^++i\varepsilon}}
\right)\vert\delta\fkj\vert^2
\end{eqnarray}
with  $E_{kj}^{\pm}\equiv E_k\pm E_j$. The account of collisional
damping is carried out in full analogy to the response function.
The $\psi_{FF}^\dpr(\om)$ is shown in Fig.\ref{fig4} by the
dotted line. Comparing with the response function (full line) the
properties just mentioned are clearly visible. In addition,
another important and generic feature is revealed. At certain
deformations the correlation function not only is finite at and
around $\om=0$ but shows a more or less pronounced and sharp peak.
For the special case shown here, which corresponds to a
deformation of $\eps =0.2$,  it appears to lie exactly at $\om=0$.
In this sense it reminds one of the "heat pole", which in previous
publications was seen to become more relevant at larger
temperatures, see \cite{hoivya} and \cite{hofrep}. However, there
are cases where such a peak is shifted to small but finite
frequencies. These peaks appear whenever a pair of levels close to
the Fermi energy approach each other; in the present case their
distance is about $\vert\epsk -\epsj\vert\approx 0.02 \mev$ which
is definitely smaller than the width associated with such
excitations.

\begin{figure}[ht]
\centerline{{\epsfxsize=16cm \epsffile{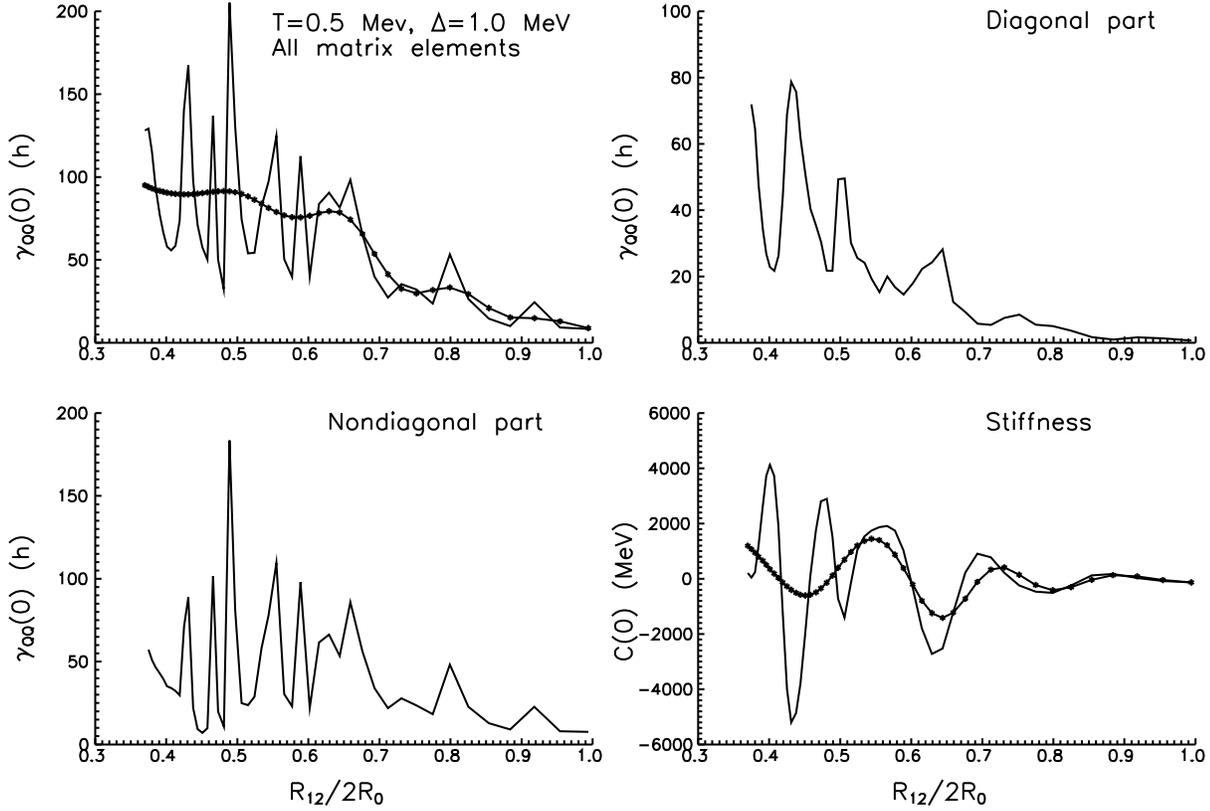}}}
\caption{The zero frequency limit \protect\req{zerofrilim} for the
friction coefficient in the shape degree of freedom $Q$ as
function of $Q$ for $T=0.5\mev$ and $\Delta=1.0\mev$. The right
bottom and top left parts of figures show separately the
contribution to the friction of nondiagonal $F_{kj}$ and diagonal
$F_{kk}$ matrix elements. The bottom right part contains the
stiffness of the static energy at $T=0.5\mev$. The curves with
stars mark the value obtained by averaging over deformation on the
interval $\Delta(R_{12}/2R_0)=0.08$. } \label{fig5}
\end{figure}

Having the correlation function at one's disposal, the friction
coefficient in zero frequency limit can be expressed as
\bel{corfric} \gamma_{FF}(0)=\psi^{\dpr}_{FF}(\om=0)/(2T)
\end{equation}
Hence, for this limit it becomes apparent how peaks in the correlation
function at $\om=0$ are related to peaks in the deformation dependence
of the friction coefficient. For the special example discussed here it
may be said that the above mentioned peak in $\psi^{\dpr}_{FF}$ at
$\om=0$ results in a peak of the friction coefficient shown in
Fig.\ref{fig5}, namely at $R_{12}/2R_0\approx 0.43$ which
corresponds to $\eps=0.2$.

Let us turn to the extreme case of $T=0$ for which $\Delta$ will
be  finite and, more important, for which the collisional width
will be zero  at least for $\hbar\om$ up to $\hbar\om=2\Delta$;
recall the discussion below eq.\req{gamma}. Look at the forms
\req{chiomlong} and \req{psiomlong} the response and correlation
functions attain for the case of no collisions. Then, only
contributions from the excitations above $2\Delta$ survive. In
case of the presence of collisions,  in the region $\om\leq
2\Delta$ both the response and correlation functions may differ
from zero only if tails from distant peaks extend into this
regime. This is very unlikely to happen if the convolution
integral in \req{chikj} is calculated exactly with the proper
frequency dependent widths. Hence, friction in zero frequency
limit will be strictly equal to zero.  However, in practical
computations where the  convolution integrals in \req{chikj} are
computed with constant $\Gamma$'s such tails might be present,
which in turn may lead to a finite though small value of the
friction coefficient. The choice used in our calculations, namely
$\Gamma=\Gamma(\Delta,\Delta,T)$, as described below \req{chikj},
does not suffer from this drawback.

As the zero frequency limit  provides the simplest definition of
friction and inertia (see \req{zerofrilim} and \req{zeromass})
respectively, we like to stick to this version for a moment. In
Fig.\ref{fig5} we exhibit the deformation dependence of
friction for the $Q$-mode, as evaluated from \req{zerofrilim}. In
this figure the rapid oscillations with deformation we spoke of
can be seen. The inertia looks similar to friction. The formulas
of the zero frequency limit \req{zerofrilim}-\req{zeromass} are
simple enough to allow one separating the contributions from
diagonal and non-diagonal parts of the response function
\req{chiomlong} (diagonal here means the contribution with $k=j$).
The source of fluctuations of these two parts turns out to be
different. The diagonal sum (contribution of the diagonal part of
the second sum in \req{chiomlong}) is affected strongly by the
density of states near the Fermi energy, which is typical of the
appearance of shell effects, of such a type as they also appear in
the static energy. They exist as long as the temperature does not
exceed a typical value of the order of $1~-~2\mev$. The stiffness
$C(0)$ of the static energy is shown in the right bottom part of
Fig.\ref{fig5}. Indeed, the same kind of fluctuations appear,
only anti-correlated to the diagonal component of friction. The
fluctuations of the non-diagonal part are caused by the
quasi-crossings of levels near the Fermi energy like those
discussed in connection to Fig.\ref{fig4}. For each sharp peak
in the left-hand-side of Fig.\ref{fig5} one can find in the
single-particle spectrum one or a few quasi-crossings which
contribute 90\% or so to friction or inertia. The height of such
peaks is the larger the smaller the minimal distance is between
quasi-crossing levels. From the analysis of the single-particle
spectrum computed with the deformed Woods-Saxon potential it
follows that some of these minimal distances are extremely small -
of the order of $0.02-0.03 MeV$.

\begin{figure}[p]
\centerline{{\epsfxsize=16cm \epsffile{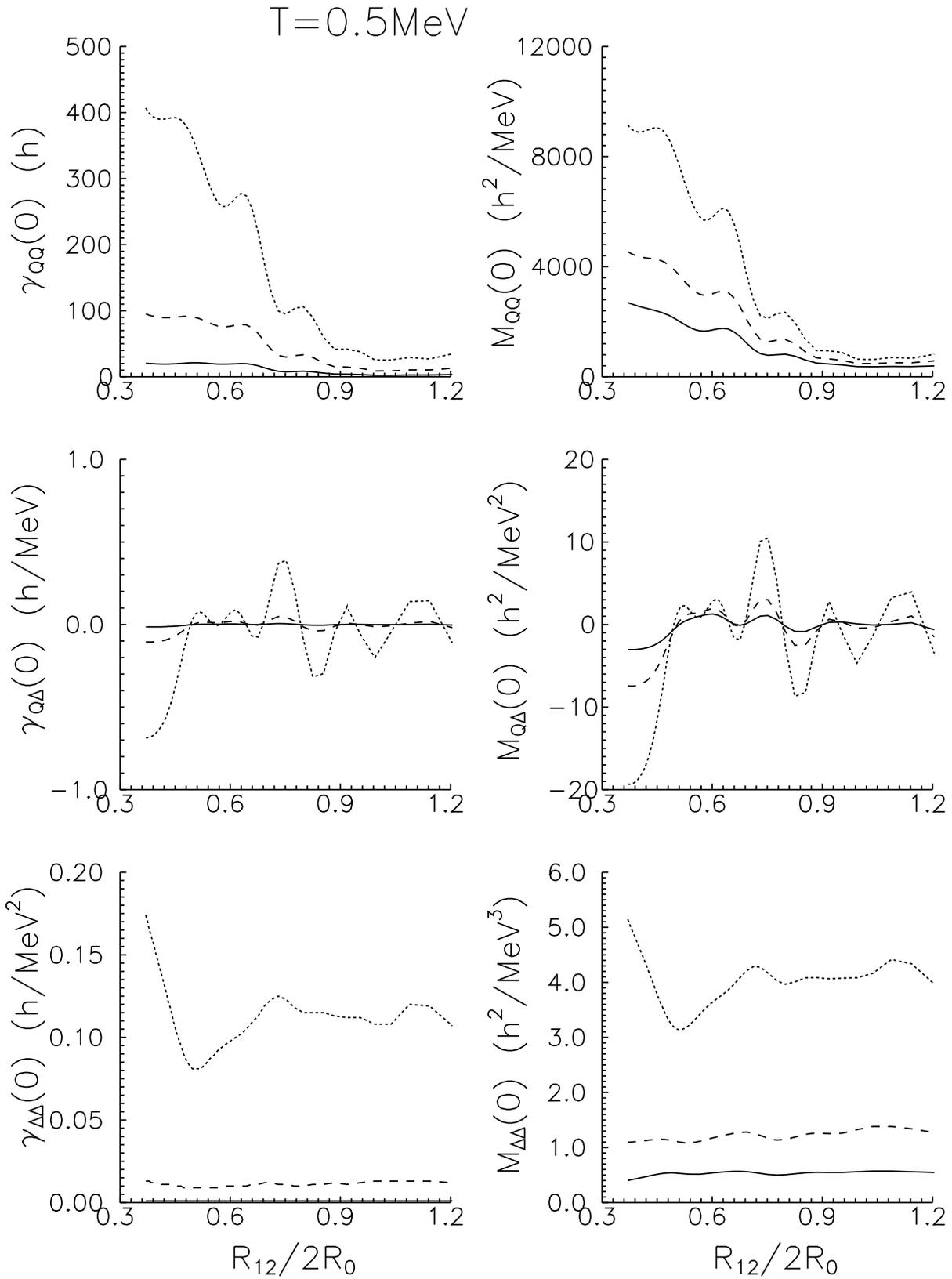}}}
\caption{The averaged values of the components of the friction
(left-hand-side) and inertial
(right-hand-side) tensors obtained within zero frequency limit
\protect\req{zerofrilim},\protect\req{zeromass} as function of
deformation for temperature $T=0.5\mev$. The dotted, dashed and solid
curves correspond to $\Delta=0.5, 1.0$ and $1.5 \mev$.}
\label{fig6}
\end{figure}

For several reasons we claim that to large extent such closely
spaced levels are shortcomings of the underlying shell model, and
that they may lead to unphysical consequences. First of all, it
should be noted that they are intimately related to the choice of
the shape parameter, in particular to the restriction to only a
few of them, more precisely to exactly one in the present case. In
a truly multi-dimensional landscape such quasi-crossings would
happen much less frequently as there is always enough room for
avoiding them. Secondly, residual interactions would reduce their
significance further. Indeed, from the statistics of
experimentally measured nuclear states it is known that levels
with small spacings are very seldom; the distribution of nearest
neighbor spacings is of Wigner type rather than of Poisson type.
Such interactions would repel the approaching states and in this
way would reduce the matrix elements with the generator $\hat F$
of collective motion. Evidently, it should be the aim of any
microscopic theory to deal with each of such problems
individually. At present, we are not yet able to do so. Rather, we
suggest to overcome these difficulties in another way, namely by
averaging the transport coefficients over deformation. The
averaging interval should be large enough to smooth out the rapid
oscillations, as well as the finer details of the shell structure,
but at the same time small enough to preserve gross shell effects.
This procedure may perhaps be motivated further by recalling the
following facts. (i) Evidently, any structure in the transport
coefficients of greater detail cannot be seen experimentally.
Usually, one is happy if one succeeds to identify gross shell
behavior. (ii) Finally, these transport coefficients are to be
used in a transport equation of Fokker-Planck or Langevin type
which account for genuine fluctuations in the collective variable.
Without any doubt, such fluctuations will smooth out the detailed
structure in the coefficients in most natural way. Incidentally,
we may phrase our problem in a different way, namely in
attributing it to a deficiency of the mean field approximation, at
least partly.


The averaged friction coefficient and stiffness are shown by the
curves with stars in Fig.\ref{fig5}. As it is seen the
frequent oscillations are gone while the typical gross structure
remains. This latter feature is clearly visible at the case of the
stiffness which now shows the typical behavior expected for a
potential with a second (and perhaps third) minimum.

In Fig.\ref{fig6} we exhibit the averaged tensors of friction
and inertia are shown, for $T=0.5 \mev$ and for a few values of
the pairing gap $\Delta$. Both friction and inertia still oscillate
as function of the deformation, but in much weaker fashion than
those shown in Fig.\ref{fig5}. Now they correlate with the
fluctuations of the stiffness or the shell correction.  The
non-diagonal terms, namely $Q\Delta$-friction and inertia, oscillate
around an average value which is very small. Very likely this
non-diagonal components of friction or inertia may simply be
neglected. The $\Delta\Delta$-friction or inertia fluctuate much less
as compared with the corresponding quantities in $Q$. The reason
is that the matrix elements of $\hat F$ (which are of strongly
peaked structure) do not contribute to $\Delta\Delta$-friction or
inertia, see \req{chipp}. The quantities which matter here most
are the single-particle energies themselves which are relatively
smooth functions of deformation. As seen from Fig.\ref{fig6},
an increase of $\Delta$ has two effects, both on friction as well as
on inertia: They become smaller and their $Q$-dependence becomes
smoother with growing $\Delta$. In case of the $QQ$-transport
coefficients the substantial contribution comes from the diagonal
component of the $\hat F$ operator (diagonal term in
eq.\req{chit}). For not too small $\Delta$ his diagonal contribution
can be estimated to behave like $\propto 1/\Delta^2$ both for
friction and inertia.  The temperature dependence (for fixed
$\Delta$) of friction and inertia is the same as in the no pairing
case: Whereas friction increases with temperature, inertia
decreases.

\subsection{Information from the collective response}
\label{subsec: numchicoll}
\begin{figure}[ht]
\centerline{{\epsfxsize=16cm \epsffile{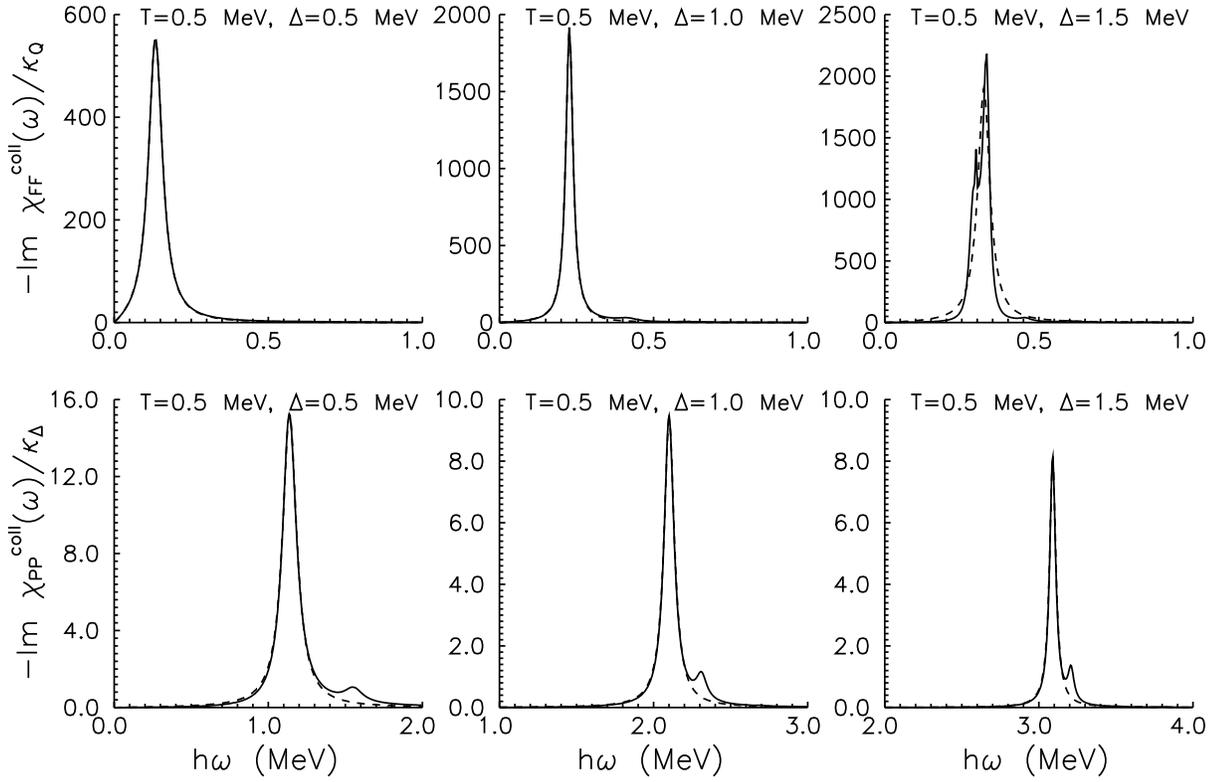}}}
\caption{The frequency dependence of the imaginary part of collective
$FF$- and $PP$- response functions, associated to the intrinsic
functions shown in Fig.\protect\ref{fig3}.  The solid and dashed
curves correspond to \protect\req{chiqqcoll},\protect\req{chippcoll} and
to the fit to \protect\req{chiqqcoll},\protect\req{chippcoll} by the
response function of a damped oscillator. }
\label{fig7}
\end{figure}

In Fig.\ref{fig7} a few examples of the tensor of the
collective response are shown for some values of $\Delta$ and
temperature. A common feature of all figures is as follows: The
collective frequency is shifted to smaller frequencies as compared
to the lowest peak of the intrinsic response. The magnitude of the
shift depends on the value of the static response
$\chi^{\pr}_{\mu\nu}(0)$  relative to the stiffness
$C_{\mu\nu}(0)$, see Fig.6 of \cite{kidhofiva} for a
demonstration. The smaller is the stiffness the larger is the shift
of collective strength to lower frequencies. In case of the
$Q$-mode the static stiffness is much smaller than
$\chi^{\pr}_{FF}(0)$ and, hence, the collective strength
concentrates at low frequencies. The situation turns out very
different for the $\Delta$-mode, whose stiffness
$C_{\Delta\Delta}$ approximately equals the static response
$\chi^{\pr}_{PP}(0)$ implying that the shift of the collective
peak is small.

\begin{figure}[ht]
\centerline{{\epsfxsize=16cm \epsffile{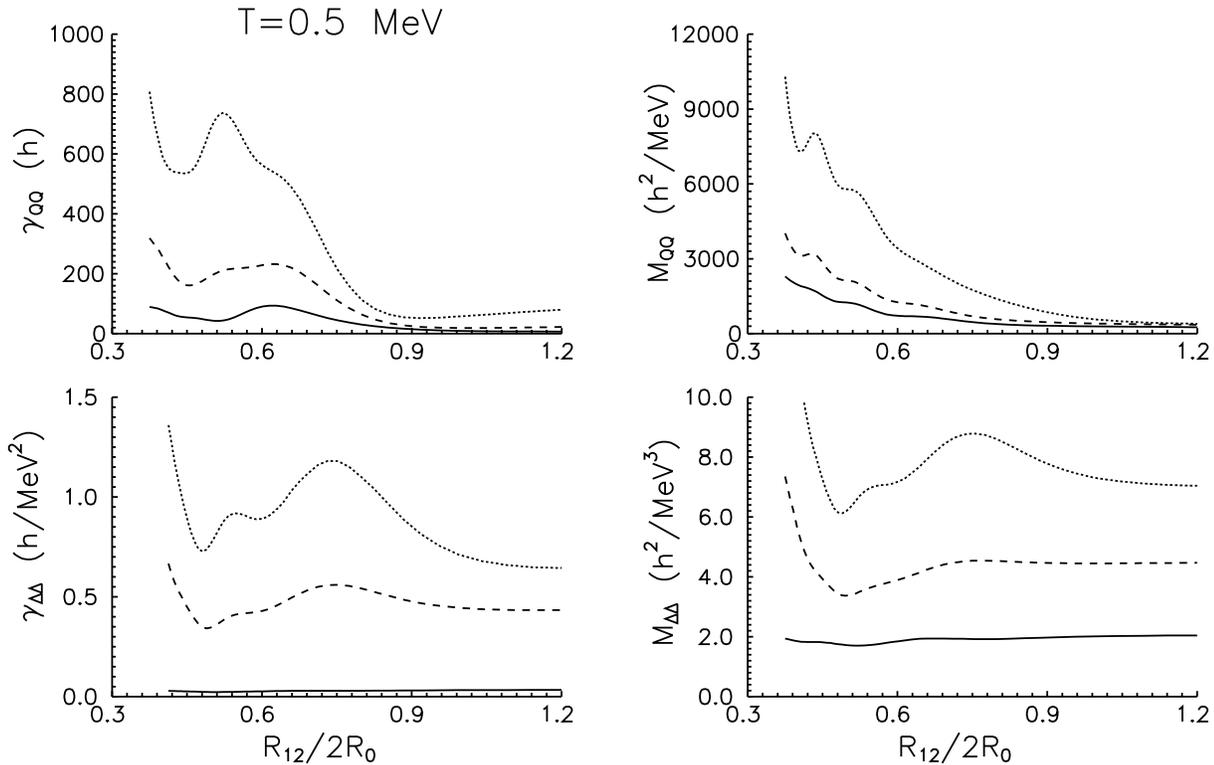}}}
\caption{The components of the friction (left-hand-side) and inertia
(right-hand-side) tensor obtained by
a fit to \protect\req{chiqqcoll},\protect\req{chippcoll} through the
response function of a damped oscillator according to \protect\req{lorfit}
as function of the deformation for temperature
$T=0.5\mev$. The dotted, dashed and solid curves correspond to $\Delta=0.5,
1.0$ and $1.5 \mev$.}
\label{fig8}
\end{figure}

In ideal cases the $\chi_{\mu\nu}^{\rm coll}(\om)$ is dominated by
one peak in the low frequency region. This peak may then be fitted
rather accurately by the response function of the damped
oscillator, in which way the self-consistent transport
coefficients are defined, see \cite{ivhopaya}. Unfortunately, in
reality one often encounters a different situation in that the
collective strength splits over several closely placed peaks which
may even overlap. This situation is typical for a
multi-dimensional case, for which it has been demonstrated in
\cite{sahoya} of how transport coefficients may still be deduced
through a generalized fitting procedure. Indeed, most likely the
present situation ought also be treated in this way.
Unfortunately, the method suggested in \cite{sahoya} is tedious
and very time consuming. In order to circumvent problems of this
type we prefer to define transport coefficients through the
approximation defined in \req{lorfit}, which still exploits the
information contained in the form $(\mbox{{\boldmath
$\kappa$}}\mbox{{\boldmath $\chi$}}_{\rm coll}~^{-1}
(\om)\mbox{{\boldmath $\kappa$}})_{\mu\nu}$ of the collective
response tensor.  The fit to the polynomial of second order given
on the right hand side of \req{lorfit} can easily be done for any
function, even if ${\mbox{{\boldmath $\chi$}}_{\rm
coll}}^{-1}(\om)$ does not look much like a polynomial. In such
cases the transport coefficients may be considered as some
frequency average over the interval of the fit. Unfortunately,
averaging in frequency does not change much the fluctuations with
respect to the deformation. That is why both quantities were also
averaged over the deformation like before. The results of such a
procedure are shown in Fig.\ref{fig8} for the diagonal
components of friction and inertia. As it is seen from the figure,
the transport coefficients obtained in this way exhibit
fluctuations with deformation similar, but noticeably different to
those of the zero frequency limit, which have been shown before in
Fig.\ref{fig6}. Moreover, the mean value of both components
of friction and inertia tensor now are larger than before. The
difference is especially large in case of $\Delta\Delta$-friction, the
reason being that  at small $\om$ the imaginary part of the
response function for the pairing degree of freedom
$\widehat\chi_{PP}(\om)$ increases with $\om$ much slower as
compared with that of a damped oscillator. Consequently,
$\gamma_{\Delta\Delta}(0)$ is smaller than $\gamma_{\Delta\Delta}$. This
is a clear hint that in such a case the zero frequency limit may not
be considered a good approximation.

\subsection{Temperature dependence of transport coefficients}
\label{subsec: tdeptran}
In the past various models have been presented to justify nuclear
dissipation, even more exist to evaluate the friction coefficient
numerically. In \cite{higoro} a compilation of data has been put
together in which theory is confronted with experimental evidence.
The latter is obtained by comparing solutions of the "macroscopic
equations" (of Fokker-Planck or Langevin type) with experimentally
observable quantities, the input into such equations being chosen
phenomenologically. It was seen that the predictions which the
theoretical models give for the effective, local collective width
$\Gamma=\gamma/M$ of the fission mode
--- sometimes referred to as the "reduced friction coefficient" $\beta$---
deviate by as much as two orders of magnitude. It can be said that
the results of \cite{yahosa} classified there by "linear response
theory" (LRT) were in good agreement with those required by
experimental evidence. As compared to the application of LRT
mentioned in \cite{higoro}, new calculations have in the mean time
been presented in \cite{ivhopaya, yaivho}. Grossly speaking their
results lie in the same regime as those of \cite{yahosa}.

As all such microscopic calculations depend on some quantities
which are not known all too precisely, it has been argued
previously (see e.g.\cite{hofrep, yaivho}) that one should look
for the temperature dependence of the effective transport
coefficients. Indeed, here the various theories differ from one
another in very pronounced ways. The wall formula, for instance,
predicts a friction coefficient which is practically independent
of $T$. As demonstrated in \cite{hoivya, hofrep}, such a picture
is obtained as the macroscopic limit of LRT, provided the latter
is applied to purely independent particle motion, viz if all
influences of collisions are discarded. Truth is that the latter
have important implications on the $T$-dependence of friction. Two
body viscosity, for instance, comes about in the regime of
"collisional dominance", in which viscosity is known to decrease
with temperature, in ideal cases like $1/T^2$. Although the notion
"viscosity" intrinsically refers to a nuclear liquid, in the case
of genuine "collisional dominance" such a $T$-dependence also
prevails for the friction coefficient of the low frequency
collective modes of finite nuclei, see e.g. \cite{maghof}. A
similar dependence was found in \cite{aynoediabfri} and
\cite{bubebr}. It is true that  these models do not rely on a
hydrodynamic description, but in each one collisions are assumed
to dominate collective dynamics in one way or another. In
\cite{hoivya} and \cite{hofrep} it has been discussed how this
behavior of damping may be obtained within LRT.

Fortunately, a possible $T$-dependence of nuclear dissipation has
also been studied by interpreting the experimental results, see
\cite{pauthoe, wisiwi, rudolf}. Here we shall not pursue any
details of these studies. It may suffice to summarize the general
claims put forward in these publications, which is that
dissipation increases with $T$ at small excitations, leveling off
or even decreasing at larger temperatures. Indeed, generally
speaking, such behaviour was found in our previous calculations
\cite{hoivya}, at least qualitatively. How large friction may
become at intermediate temperature and how quickly it decreases
afterwards is still an unsettled question. It depends both on the
contributions from the heat pole as well as on the approximations
in which collisions are treated, for details see \cite{hoivya} and
\cite{hofrep}. As stressed earlier, in the present paper we mainly
address the regime of very small temperatures where pairing must
not be neglected.  Unfortunately, to the best of our knowledge, in
this regime, neither experimental information is available, nor
to date is there any theoretical model with which our results
might be compared. Our predictions are as follows.

\begin{figure}[t]
\centerline{{\epsfxsize=16cm \epsffile{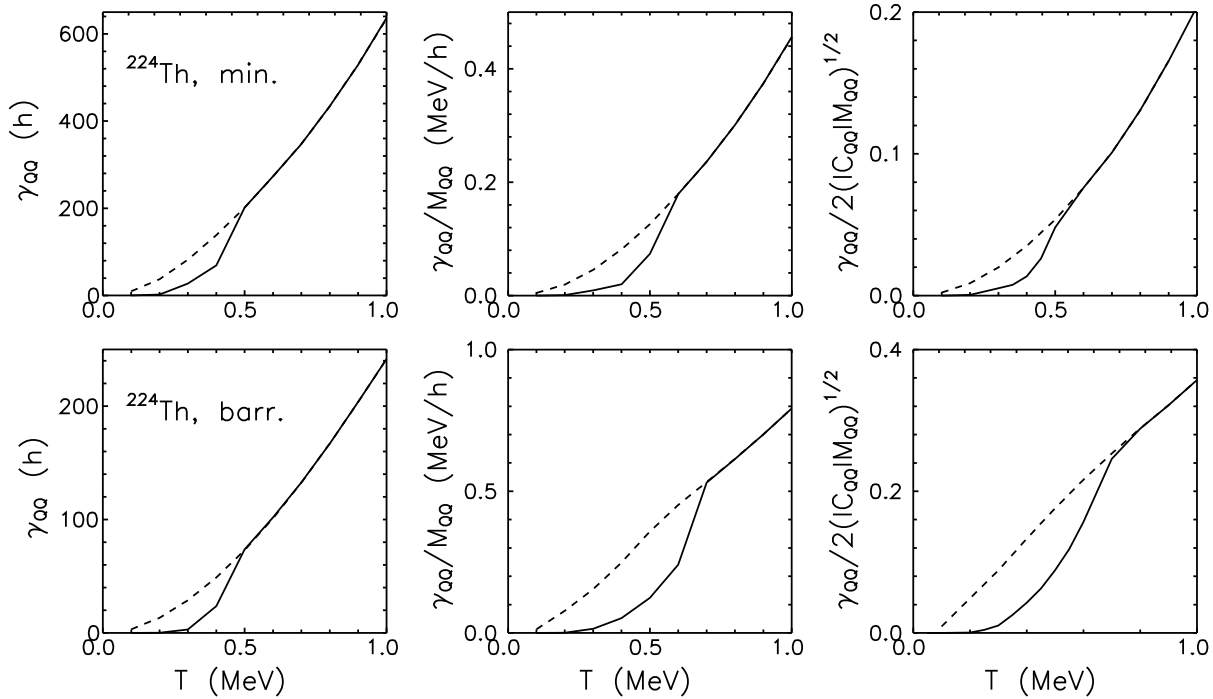}}} \caption{The
average $QQ$-friction coefficient $\gamma_{QQ}$, reduced friction
coefficient $\beta_{QQ}\equiv\gamma_{QQ}/M_{QQ}$ and the damping
factor $\eta_{QQ}= \gamma_{QQ}/2\protect\sqrt{C_{QQ}M_{QQ}}$ at
the ground state (top) and fission barrier (bottom) of $^{224}Th$
as function of temperature.  The dashed curves show the results
obtained neglecting pairing.  } \label{fig9}
\end{figure}

The ones for the fission mode are displayed in Fig.\ref{fig9}.
To simplify matters the pairing mode has been treated in
equilibrium, or "local" equilibrium, rather, if the latter is
being defined with respect to the collective variable $Q$. In
other words, at given shape the gap equation \req{gengapeq} has
been solved for various temperatures. The resulting $\Delta$'s
then have been used for the calculation of the transport
coefficients for the $Q$-mode displayed in Fig.\ref{fig9}. For the
system considered pair correlations disappear above  $T\simeq
0.5\mev$. Indeed, it is only below such temperatures that the
transport coefficients deviate from the unpaired case. At first
sight the deviation may look small. A closer look, however, shows
that important modifications of the nature of the transport
process are present. Neglecting any quantum corrections (for
collective dynamics) one realizes that for smaller temperatures
nuclear dissipation may get so weak that one reaches Kramers' so
called "low viscosity" limit". This is very interesting on
recalling that in nuclear physics experimental results are
commonly interpreted on the basis of Kramers' simple formula for
the "high viscosity limit". Moreover, it so happens that in this
regime quantum effects in collective dynamics are present. They
lead to additional modifications of the decay rate. Unfortunately,
at present one is able to account for these effects only above
some critical temperature, which turns out to be less than the $T$
where pairing disappears. There is no room to elaborate further on
details of collective dynamics. Instead we like to refer to a
recent Letter \cite{hifed}, as well as to forthcoming papers.

\begin{figure}[t]
\centerline{{\epsfxsize=10cm \epsffile{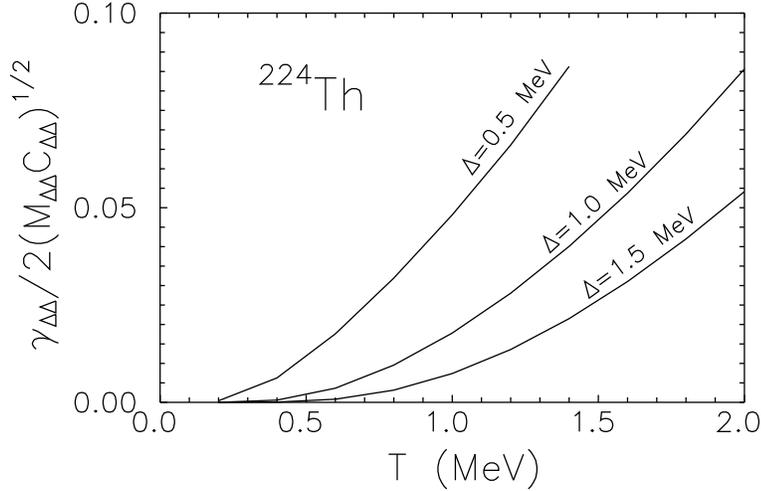}}} \caption{The
damping parameter of pairing vibrations
$\protect\bar\eta_{\Delta\Delta}= \gamma_{\Delta\Delta}/
2\protect\sqrt{C_{\Delta\Delta}M_{\Delta\Delta}}$ as the function
of temperature for few fixed values of $\Delta$. The damping
parameter is averaged along the fission valley of nucleus
$^{224}Th$. } \label{fig10}
\end{figure}

Let us turn to the pairing mode $\Delta$ now, for which in
Fig.\ref{fig10} we show the damping factor $\eta_{\Delta\Delta}=
\gamma_{\Delta\Delta}/ 2\sqrt{C_{\Delta\Delta}M_{\Delta\Delta}}$
as function of temperature for various values of $\Delta$. Here,
the latter is to be understood as a collective variable, not fixed
by the gap equation. The values presented in this figure represent
averages along the fission valley. The $\eta_{\Delta\Delta}$ is
seen to be very small, implying the corresponding pairing
vibrations to be strongly {\it under}-damped. Most likely this
means that on the way from saddle to scission the pairing mode
does {\it not equilibrize}. The consequences such an observation
may have on odd-even effects in mass distributions etc. will have
to be the subject of some further studies.

\section{Summary}
\label{sec: summ}

A detailed study of transport coefficients has presented for
temperatures below $T=1 MeV$ which finally aim at a description of
the dynamics of fissioning nuclei. Pairing effects were accounted
for within the independent quasi-particle approximation. An
appropriate treatment of particle number conservation has been
achieved by modifying the response functions of the locally
harmonic approximation. For harmonic vibrations without damping it
has been demonstrated that the same secular equation for the
collective frequencies is obtained as those derived earlier within
common RPA. In our version, the pairing gap $\Delta$ is introduced
as an independent dynamical variable, similar to the parameters
specifying the shape of nuclear surface. The response functions
and the tensors of collective friction and inertia have been
calculated for a realistic, deformed Woods-Saxon potential. The
components of friction and inertia tensors were examined as
function of the pairing gap and the deformation of the nucleus.
With respect to the latter strong variations were found of all
transport coefficients. They are caused both by shell effects as
well as by avoided crossings of single-particle levels. It has
been argued that such variations may be considered as carrying
unphysical features, for which reason they have been smoothed out
by averaging over the shape in such a way as to keep gross shell
effects. As could be expected, the friction coefficient decreases
with increasing $\Delta$. Finally we may say that the temperature
dependence of the friction coefficient obtained  in our model is
in qualitative agreement with the conclusions reached in
\cite{pauthoe, wisiwi, rudolf}, namely that dissipation increases
with $T$ at small excitations.

\bigskip

{\bf Acknowledgments}. The authors want to thank the Deutsche
Forschungsgemeinschaft for financial support and to D. Einzel,
A.S. Jensen, K. Hara, K. Pomorski and V. Abrosimov for fruitful
discussions and suggestions. The use of V.V. Pashkevich's
computational code for the quantities based on the shell model
with deformed Woods-Saxon potential is greatly acknowledged.  One
of us (F.A.I) would like to thank the Physik Department of the TUM
for the hospitality extended to him during his stay in Garching.

\appendix
\section{Intrinsic response functions}
\label{sec: respfuns}
The time dependent response function $\tilde\chi_{FF}(t)$ is defined by
\req{defrest}. To obtain $\tilde\chi_{PN}(t)$ one should replace
$\foper^I(t),\foper^I(s)$ in \req{defrest} by $(\hat P+\hat P^\dagger)^I(t),
\hat N^I(s)$. In the same way one can obtain expressions for the other
response functions, too. For the operators $\hat P$ and $\hat N$ one may
use the quasiparticle representation like \req{operator}
\begin{eqnarray}
\label{operators}
\hat P+\hat P^\dagger=\sum_k 2u_k v_k -\sum_k 2u_k v_k(\alkp\alk+\alkbp\alkb)
+\sum_k(u_k^2-v_k^2)(\alkp\alkbp+\alkb\alk)\nonumber\\
\hat N=\sum_k 2v_k^2 +\sum_k(u_k^2-v_k^2)(\alkp\alk+\alkbp\alkb)
+\sum_k 2u_k v_k (\alkp\alkbp+\alkb\alk)\nonumber\\
\end{eqnarray}
The final expressions for $FP$ or $FN$-response functions turn out to
be much simpler than the one presented in \req{chiom}  since they
contain only diagonal sums $(k=j)$.
\begin{eqnarray}\label{respfuns}
\chi_{FN}(\om)&=&\sum_k (2n_k^T-1)(2u_k v_k)^2 F_{kk}
\left({1\over{\hbar\om-2E_k+i\varepsilon}}-{1\over{\hbar\om+2E_k+i\varepsilon}}
\right)\nonumber\\
\chi_{FP}(\om)&=&\sum_k (2n_k^T-1)2u_k v_k (u_k^2-v_k^2)F_{kk}
\left({1\over{\hbar\om-2E_k+i\varepsilon}}-{1\over{\hbar\om+2E_k+i\varepsilon}}
\right)\nonumber\\
\chi_{NN}(\om)&=&\sum_k (2n_k^T-1)(2u_k v_k)^2
\left({1\over{\hbar\om-2E_k+i\varepsilon}}-{1\over{\hbar\om+2E_k+i\varepsilon}}
\right)\nonumber\\
\chi_{NP}(\om)&=&\sum_k (2n_k^T-1)2u_k v_k (u_k^2-v_k^2)
\left({1\over{\hbar\om-2E_k+i\varepsilon}}-{1\over{\hbar\om+2E_k+i\varepsilon}}
\right)\nonumber\\
\chi_{PP}(\om)&=&\sum_k(2n_k^T-1)(u_k^2-v_k^2)^2
\left({1\over{\hbar\om-2E_k+i\varepsilon}}-{1\over{\hbar\om+2E_k+i\varepsilon}}
\right)\nonumber\\
\end{eqnarray}
The account for the collisional damping for $FP$- $FN$- and other
response function is carried out in the same way as for $FF$-response
and leads to
\begin{eqnarray}\label{damprespfuns}
\chi_{FN}(\om)&=&\sum_k (2u_k v_k)^2 F_{kk}
\sum_{s\neq s^{\pr}}\chi_{ksks^{\pr}}(\om) \nonumber\\
\chi_{FP}(\om)&=&\sum_k 2u_k v_k (u_k^2-v_k^2)F_{kk}
\sum_{s\neq s^{\pr}}\chi_{ksks^{\pr}}(\om)\nonumber\\
\chi_{NN}(\om)&=&\sum_k (2u_k v_k)^2
\sum_{s\neq s^{\pr}}\chi_{ksks^{\pr}}(\om)\nonumber\\
\chi_{NP}(\om)&=&\sum_k 2u_k v_k (u_k^2-v_k^2)
\sum_{s\neq s^{\pr}}\chi_{ksks^{\pr}}(\om)\nonumber\\
\chi_{PP}(\om)&=&\sum_k (u_k^2-v_k^2)^2
\sum_{s\neq s^{\pr}}\chi_{ksks^{\pr}}(\om)\nonumber\\
\end{eqnarray}
with $\chi_{ksks^{\pr}}(\om)$ given by \req{chikj}

\end{document}